\documentclass{article}
\usepackage{mathbbold}
\usepackage{amsfonts}
\usepackage{color}
\usepackage{float}
\usepackage{cite}
\usepackage{graphicx}
\usepackage{amsmath}

\def\R{{\mathbb{R}}}

\def\C{\mathbb{C}}

\def\OO{{\otimes}}

\def\H{{\mathcal{H}}}
\date{\empty}
\def\<{\langle}
\def\>{\rangle}
\def\tr{{\rm{tr}}}
\def\lm{\lambda}
\newcommand{\be}{\begin{equation}}
\newcommand{\ee}{\end{equation}}
\newcommand{\bea}{\begin{eqnarray}}
\newcommand{\eea}{\end{eqnarray}}
\newcommand{\bean}{\begin{eqnarray*}}
\newcommand{\eean}{\end{eqnarray*}}
\def\Bq{\begin{equation}\label}
\def\Eq{\end{equation}}\renewcommand{\theequation}{\thesection.\arabic{equation}}

\begin{document}

\title{\bf Characterizations of bilocality and $n$-locality of  correlation  tensors}
\author{Shu Xiao, Huaixin Cao, Zhihua Guo,  Kanyuan Han\\
{\it\small School of Mathematics and Statistics, Shaanxi Normal University}\\
{\it\small Xi'an 710119, China}}
\maketitle
\begin{abstract}In the literature, bilocality and $n$-locality of correlation tensors (CTs) are described by integration local hidden variable models (called C-LHVMs) rather than by summation LHVMs (called D-LHVMs). Obviously, C-LHVMs are easier to be constructed than D-LHVMs, while the later are easier to be used than the former, e.g., in discussing on the topological and geometric properties of the sets of all bilocal  and  of all $n$-local CTs. In this context,
one may ask whether the two descriptions are equivalent. In the present work, we first establish some equivalent characterizations of  bilocality of a tripartite CT ${\bf{P}}=\Lbrack P(abc|xyz)\Rbrack$, implying that the two descriptions of bilocality are equivalent. As applications, we prove that all bilocal CTs with the same size form a compact path-connected set that has many star-convex subsets. Secondly, we  introduce and discuss the bilocality of a tripartite probability tensor (PT) ${\bf{P}}=\Lbrack P(abc)\Rbrack$, including  equivalent characterizations and properties of bilocal PTs. Lastly, we obtain corresponding  results about $n$-locality of $n+1$-partite CTs ${\bf{P}}=\Lbrack P({\bf{a}}b|{\bf{x}}y)\Rbrack$  and PTs ${\bf{P}}=\Lbrack P({\bf{a}}b)\Rbrack$, respectively.

\end{abstract}

{\bf PACS number(s)}: 03.65.Ud, 03.67.Mn

\section{Introduction}

As one of important quantum correlations, Bell nonlocality originated from  the Bell's 1964 paper \cite{Bell}. He found that when some entangled state is suitably measured, the probabilities for the outcomes violate an inequality,  named the Bell inequality. This property of quantum states is the so-called Bell nonlocality  and was reviewed by Brunner et al. \cite{Brunner} for the ``behaviors" $P(ab|xy)$ (correlations), a terminology introduced by Tsirelson \cite{Tsir}, but not for quantum states. As an important source in quantum information processing, Bell nonlocality has been widely discussed, see e.g.  \cite{[1], Fine, Werner1989,  [2], Werner, Col, [6], Barrett2002, JLChen, Wies, Barrett2005, Wiseman, Sujit, Buhrman,CaoGuo}.
Usually, Bell nonlocality can be checked by violation of some types Bell inequalities \cite{Gisin1, Khal, Gisin2, Ard, Bel, Horodecki, And, T, YuSX,LiMing,Hoban}.

\begin{figure}[h]
  \centering
  \includegraphics[width=6cm]{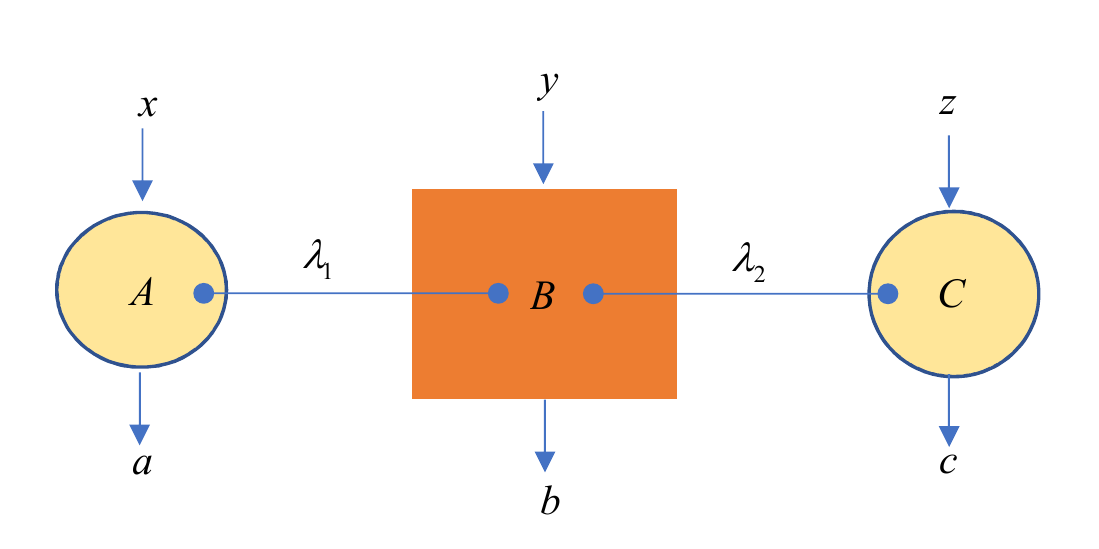}\\
  \caption{A general bilocal scenario. A source $\lim_1$ sends particles to Alice and Bob, and a separate source $\lm_2$ sends particles to Charles and Bob. All parties can perform measurements on their system,
labeled $x, y$, and $z$ for Alice, Bob, and Charles, and they obtain outcomes $a, b$, and $c$, respectively, with the joint probability $P(abc|xyz)$.}\label{Bilocal1}
\end{figure}

Quantum systems that have never interacted can become nonlocally correlated through a process called entanglement swapping. To characterize nonlocality in this context, Branciard et al. \cite{Branciard2010} introduced local models where quantum systems that are initially uncorrelated are described by uncorrelated local variables, leading to stronger tests of nonlocality.   More precisely, they considered the general scenario depicted in Fig. \ref{Bilocal1}.

\begin{figure}[h]
  \centering
  \includegraphics[width=7cm]{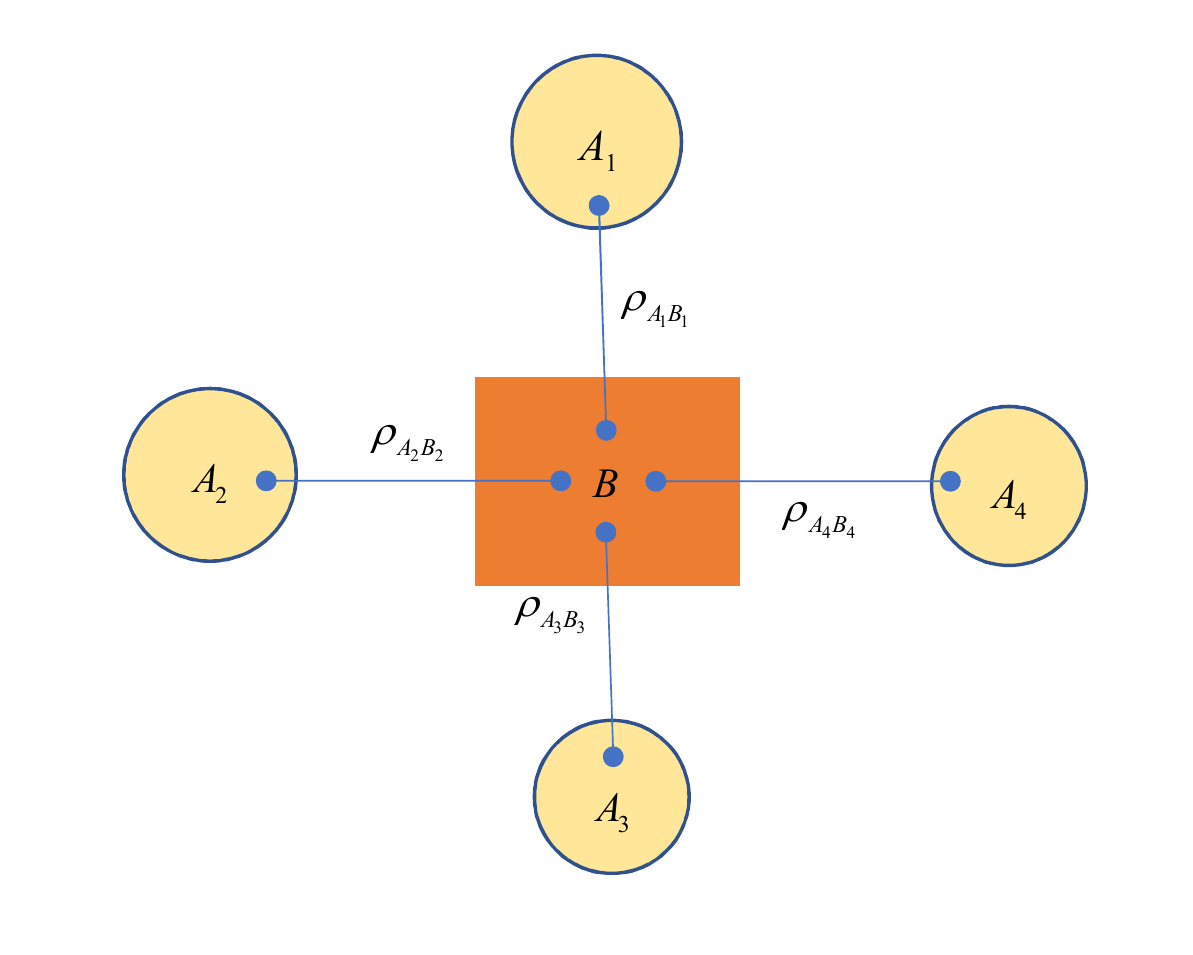}\\
  \caption{A star-network with a central node $B$ and $n$ star-nodes $A_1,A_2,\ldots,A_n$ where $n=4$.}\label{starQN}
\end{figure}

After performing measurements, the correlations between the measurement outcomes of the three parties are described by the joint probability distribution $P(abc|xyz)$. Following \cite{Branciard2010,Branciard2012},  a joint probability distribution is said to be {\it bilocal} if it can be
written in the factorized form:
{\begin{equation}\label{bi-LC}
P(abc|xyz)=\iint_{\Lambda_1\times\Lambda_2}\rho_1(\lm_1)\rho_2(\lm_2)
P_A(a|x,\lm_1)P_B(b|y,\lm_1\lm_2)P_C(c|z,\lm_2){\rm{d}}\lm_1{\rm{d}}\lm_2
\end{equation}}
for all possible inputs $x,y,z$ and all outcomes $a,b,c$. We call Eq. (\ref{bi-LC}) a continuous bilocal hidden variable model (C-biLHVM) since hidden variables $\lm_1$ and $\lm_2$ may be ``continuous" ones. Branciard et al. proved that all bilocal correlations satisfy a quadratic inequality $I\le 1+E^2$
\cite[Eq. (10)]{Branciard2010}.
To compare bilocal and nonbilocal correlations in entanglement-swapping experiments, Branciard et al. \cite{Branciard2012} extended   the analysis of bilocal correlations initiated in \cite{Branciard2012} and
derived a Bell-type inequality $\sqrt{I}+\sqrt{J}\le1,$ which was proved to be valid for every bilocal ${\bf{P}}=\Lbrack P(abc|xyz)\Rbrack$. Gisin et al. \cite{Gisin3} proved that all entangled pure quantum states violate the bilocality inequality.
Importantly, bilocality inequality is related to the
2-locality approach for detecting quantum correlations in
networks, especially, in star networks  \cite{[21], [23],Tavakoli,LuoMX,Fran, Kri,Gisin2017, Gisin2019,Mukherjee,YC,Renou}. For example, Tavakoli et al in \cite{[21]} introduced and discussed $n$-locality of a star-network  composed by $n+1$ parties (see Fig. \ref{starQN}), where a central node
(referred to as Bob, denoted by $B$) shares a bipartite state $\rho_{A_iB_i}$ with each $A_i$ of $n$ edge nodes (referred to as the Alices, denoted by $A_1,A_2,\ldots,A_n$).

Usually, both bilocality and $n$-locality are described by integration local hidden variable models (LHVMs), called C-biLHVMs and C-$n$LHVMs, rather than by summations LHVMs, called  D-biLHVMs and D-$n$LHVMs.
In this paper, we will discuss bilocality and $n$-locality of  correlation and probability tensors by proving equivalences between C-biLHVMs and  D-biLHVMs, as well as C-$n$LHVMs and D-$n$LHVMs. In Sect. 2, we first fix the concept of bilocality of a tripartite correlation tensor (CT) ${\bf{P}}=\Lbrack P(abc|xyz)\Rbrack$, and establish  a series of characterizations and many properties of bilocality. In Sect. 3, we give the concept of bilocality of a probability tensor (PT) ${\bf{P}}=\Lbrack P(a,b,c)\Rbrack$ and obtain some equivalent characterizations and many properties of bilocality of a PT. Sects. 4 and 5 are devoted to the corresponding discussions about $n+1$-partite CTs ${\bf{P}}=\Lbrack P({\bf{a}}b|{\bf{x}}y)\Rbrack$  and PTs ${\bf{P}}=\Lbrack P({\bf{a}}b)\Rbrack$, respectively.

\section{Bilocality of tripartite correlation tensors}
\setcounter{equation}{0}

In what follows, we use $[n]$ to denote the set $\{1,2,\ldots,n\}$. When  a tripartite system is measured by separated three parties $A,B$ and $C$ with measurements labeled by $x\in[m_A],y\in[m_B]$ and $z\in[m_C]$, respectively, the joint probability distribution $P(abc|xyz)$ of obtaining outcomes $a\in[o_A],b\in[o_B]$ and $c\in[o_C]$ forms a tensor ${\bf{P}}=\Lbrack P(abc|xyz)\Rbrack$  over $\Delta_3=[o_A]\times[o_B]\times[o_C]\times[m_A]\times[m_B]\times[m_C]$, we call it a {\it correlation tensor} (CT) \cite{BaiLH}, just like a matrix. Abstractly, a tripartite CT over $\Delta_3$ is a function $P:\Delta_3\rightarrow\R$ such that
$$P(abc|xyz)\ge0 (\forall x,y,z,a,b,c){\textrm{\ and\ }} \sum_{a,b,c}{P(abc|xyz)}=1 (\forall x,y,z).$$
Any function $P:\Delta_3\rightarrow\R$ is called a {\it correlation-type tensor} (CTT)\cite{BaiLH} over $\Delta_3$.
We use $\mathcal{T}(\Delta_3)$ and $\mathcal{CT}(\Delta_3)$ to denote the sets of all CTTs and CTs over $\Delta_3$, respectively.

For any two elements ${\bf{P}}=\Lbrack P(abc|xyz)\Rbrack$ and  ${\bf{Q}}=\Lbrack Q(abc|xyz)\Rbrack$ of $\mathcal{T}(\Delta_3)$, define
$$ s{\bf{P}}+t{\bf{Q}}=\Lbrack sP(abc|xyz)+tQ(abc|xyz)\Rbrack,$$$$
\<{\bf{P}}|{\bf{Q}}\>=\sum_{a,b,c,x,y,z}{P(abc|xyz)}Q(abc|xyz),$$
then $\mathcal{T}(\Delta_3)$ becomes a finite dimensional Hilbert space over $\R$.
Clearly, the norm-convergence of a sequence in $\mathcal{T}(\Delta_3)$ is just the pointwise-convergence and then $\mathcal{CT}(\Delta_3)$ becomes a compact convex set in $\mathcal{T}(\Delta_3)$.

We fix the concept of the bilocality of a CT  over $\Delta_3$ according to \cite{Branciard2010,Branciard2012}.

{\bf Definition 2.1.} A CT   ${\bf{P}}=\Lbrack P(abc|xyz)\Rbrack$  over $\Delta_3$ is said to be {\it bilocal} if it  has a ``continuous" bilocal hidden variable model (C-biLHVM):
{\footnotesize\begin{equation}\label{bi-MCT22}
P(abc|xyz)=\iint_{\Lambda_1\times\Lambda_2}q_1(\lm_1)q_2(\lm_2)
P_A(a|x,\lm_1)P_B(b|y,\lm_1\lm_2)P_C(c|z,\lm_2){\rm{d}}\mu_1(\lm_1){\rm{d}}\mu_2(\lm_2)
\end{equation}
}
for a measure space $(\Lambda_1\times\Lambda_2,\Omega_1\times\Omega_2,\mu_1\times\mu_2)$ and for all $a,b,c,x,y,z$, where

(a) $q_1(\lm_1)$ and $P_A(a|x,\lm_1)(x\in[m_A],a\in[o_A])$ are  $\Omega_1$-measurable on $\Lambda_1$,  $q_2(\lm_2)$ and $P_C(c|z,\lm_2)(z\in[m_C],c\in[o_C])$ are  $\Omega_2$-measurable on $\Lambda_2$, and $P_B(b|y,\lm_1\lm_2)(y\in[m_B],b\in[o_B])$ are  $\Omega_1\times\Omega_2$-measurable on $\Lambda_1\times\Lambda_2$;

(b) $q_i(\lm_i),P_A(a|x,\lm_1), P_B(b|y,\lm_1\lm_2)$ and $P_C(c|z,\lm_2)$ are probability distributions (PDs) of $\lm_i,a,b,c,$ respectively.

A CT   ${\bf{P}}=\Lbrack P(abc|xyz)\Rbrack$  over $\Delta_3$ is said to be {\it non-bilocal} if it is not bilocal.  We use  $\mathcal{CT}^{\textrm{bilocal}}(\Delta_3)$ to denote the set of all bilocal CTs over $\Delta_3$.

{\bf Remark 2.1.} By Definition 2.1, when a CT ${\bf{P}}=\Lbrack P(abc|xyz)\Rbrack$ over $\Delta_3$ is a product of three conditional probability distributions $P_A(a|x),P_B(b|y)$ and $P_C(c|z)$ of parties $A,B$ and $C$,  i.e., $P(abc|xyz)=
P_A(a|x)P_B(b|y)P_C(c|z)$, we can rewrite it as
$$P(abc|xyz)=\sum_{\lm_1\lm_2=1}^1q_1(\lm_1)q_2(\lm_2)P_A(a|x,\lm_1)P_B(b|y,\lm_1\lm_2)P_C(c|z,\lm_2)$$
where  $q_k(\lm_k)=1(k=1,2)$ and $$P_A(a|x,\lm_1)=P_A(a|x),P_B(b|y,\lm_1\lm_2)=P_B(b|y),P_C(c|z,\lm_2)=P_C(c|z),\ \forall\lm_k=1.$$
Thus, ${\bf{P}}=\Lbrack P(abc|xyz)\Rbrack$ can be written as (\ref{bi-MCT22}) for the counting measures $\mu_k$ on the set $\Lambda_k=\{1\}$ and then is bilocal.

{\bf Remark 2.2.} From definition, we observe that when  a CT ${\bf{P}}=\Lbrack P(abc|xyz)\Rbrack$  over $\Delta_3$ is  bilocal, marginal distributions satisfy:
\Bq{Nec}P_{AC}(ac|xz)=P_{A}(a|x)P_{C}(c|z),\ \ \forall x,z,a,c.\Eq
By using this property of a bilocal CT, we can find that not all Bell local CTs over   $\Delta_3$ are  bilocal.

{\bf Example 2.1.} Let $m_X=o_X=2(X=A,B,C)$ and take
$$P_B(1|1)=1/2,P_B(2|1)=1/2,P_B(1|2)=1/2,P_B(2|2)=1/2,$$
$$P'_A(1|1)=1,P'_A(2|1)=0,P'_A(1|2)=1,P'_A(2|2)=0,$$$$
P''_A(1|1)=0,P''_A(2|1)=1,P''_A(1|2)=0,P''_A(2|2)=1,$$
$$P'_C(1|1)=1,P'_C(2|1)=0,P'_C(1|2)=0,P'_C(2|2)=1,$$$$
P''_C(1|1)=0,P''_C(2|1)=1,P''_C(1|2)=1,P''_C(2|2)=0,$$
and define $${\bf{P}}=\frac{1}{2}{\bf{P}}'_A\OO{\bf{P}}_B\OO{\bf{P}}'_C
+\frac{1}{2}{\bf{P}}''_A\OO{\bf{P}}_B\OO{\bf{P}}''_C,$$
that is,
$$P(abc|xyz)=\frac{1}{2}P'_A(a|x)P_B(b|y)P'_C(c|z)
+\frac{1}{2}P''_A(a|x)P_B(b|y)P''_C(c|z).$$
Clearly, ${\bf{P}}=\Lbrack P(abc|xyz) \Rbrack$ is a Bell local CT. Note that
$$|P_{AC}(ac|xz)-P_{A}(a|x)P_{C}(c|z)|=\frac{1}{4}|[
P'_A(a|x)-P''_A(a|x)][
P'_C(c|z)-P''_C(c|z)]|\equiv\frac{1}{4},$$
we see that
$$P_{AC}(ac|xz)\ne P_{A}(a|x)P_{C}(c|z),\ \ \forall x,z,a,c.$$
Thus, ${\bf{P}}\notin\mathcal{CT}^{\textrm{bilocal}}(\Delta_3)$. Moreover,
from Remark 2.1, we see that ${\bf{P}}'_A\OO{\bf{P}}_B\OO{\bf{P}}'_C$ and ${\bf{P}}''_A\OO{\bf{P}}_B\OO{\bf{P}}''_C$ are in $\mathcal{CT}^{\textrm{bilocal}}(\Delta_3)$. This shows that the set $\mathcal{CT}^{\textrm{bilocal}}(\Delta_3)$ is not convex in the Hilbert space $\mathcal{T}(\Delta_3)$.

Generally, the five PDs in a C-biLHVM (\ref{bi-MCT22}) for ${\bf{P}}$ are not necessarily unique and depend on ${\bf{P}}$. The following proposition ensures that any two bilocal CTs over $\Delta_3$ can be represented by C-biLHVMs with the same measure space and the same PDs of the same hidden variables.

{\bf Proposition 2.1.} {\it Let ${\bf{P}}=\Lbrack P(abc|xyz)\Rbrack$ and ${\bf{P}}'=\Lbrack P'(abc|xyz)\Rbrack$ be any two bilocal CTs over $\Delta_3$. Then there is a product measure space $(S_1\times S_2,T_1\times T_2,\gamma_1\times\gamma_2)$ and PDs $f_k(s_k)$ of $s_k(k=1,2)$  such that}
{\begin{eqnarray}\label{bi-p}
P(abc|xyz)&=&\iint_{S_1\times S_2}f_1(s_1)f_2(s_2)
P_A(a|x,s_1)\nonumber\\
&&\times P_B(b|y,s_1s_2)P_C(c|z,s_2){\rm{d}}\gamma_1(s_1){\rm{d}}\gamma_2(s_2),
\end{eqnarray}
\begin{eqnarray}\label{bi-p'}
P'(abc|xyz)&=&\iint_{S_1\times S_2}f_1(s_1)f_2(s_2)
P'_A(a|x,s_1)\nonumber\\
&&\times P_B'(b|y,s_1s_2)P_C'(c|z,s_2){\rm{d}}\gamma_1(s_1){\rm{d}}\gamma_2(s_2)
\end{eqnarray}
for all $a,b,c,x,y,z.$}

{\bf Proof.} By definition, ${\bf{P}}$ and ${\bf{P}}'$ can be represented as
{\begin{eqnarray}\label{b12}
P(abc|xyz)&=&\iint_{\Lambda_1\times\Lambda_2}q_1(\lm_1)q_2(\lm_2)
P_A(a|x,\lm_1)\nonumber\\
&&\times P_B(b|y,\lm_1\lm_2)P_C(c|z,\lm_2){\rm{d}}\mu_1(\lm_1){\rm{d}}\mu_2(\lm_2)
\end{eqnarray}}
{\begin{eqnarray}\label{b22}
P'(abc|xyz)&=&\iint_{\Lambda_1'\times\Lambda_2'}q_1'(\lm_1')q_2'(\lm_2')
P_A'(a|x,\lm_1')\nonumber\\
&&\times P_B'(b|y,\lm_1'\lm_2')P_C'(c|z,\lm_2'){\rm{d}}\mu_1'(\lm_1')
{\rm{d}}\mu_2'(\lm_2'),
\end{eqnarray}}
for some product measure spaces  $(\Lambda_1\times\Lambda_2,\Omega_1\times\Omega_2,\mu_1\times\mu_2)$
and $(\Lambda_1'\times\Lambda_2',\Omega_1'\times\Omega_2',\mu_1'\times\mu_2').$ Putting
$$S_k=\Lambda_k\times\Lambda_k',T_k=\Omega_k\times\Omega_k',\gamma_k=\mu_k\times\mu_k',$$
$$s_k=(\lm_k,\lm_k'),f_k(s_k)=q_k(\lm_k)q_k'(\lm_k')$$
for $k=1,2$, we get a product measure space $(S_1\times S_2,T_1\times T_2,\gamma_1\times\gamma_2)$, two independent variables $s_k$ with distributions $f_k(s_k)(k=1,2)$. By letting
$$P_A(a|x,s_1)=P_A(a|x,\lm_1),
P_B(b|y,s_1s_2)=P_B(b|y,\lm_1\lm_2), P_C(c|z,s_2)=P_C(c|z,\lm_2),$$
$$P_A'(a|x,s_1)=P_A'(a|x,\lm_1'),
P_B'(b|y,s_1s_2)=P_B'(b|y,\lm_1'\lm_2'), P_C'(c|z,s_2)=P_C'(c|z,\lm_2'),$$
for all $s_k=(\lm_k,\lm_k')$ in $S_k$, we obtain from  Eq. (\ref{b12}) that
\begin{eqnarray*}
&&\iint_{S_1\times S_2}f_1(s_1)f_2(s_2)
P_A(a|x,s_1)P_B(b|y,s_1s_2)P_C(c|z,s_2){\rm{d}}\gamma_1(s_1){\rm{d}}\gamma_2(s_2)\\
&=&\iint_{\Lambda_1'\times \Lambda_2'}q_1'(\lm_1')q_2'(\lm_2'){\rm{d}}\mu_1'(\lm_1')
{\rm{d}}\mu_2'(\lm_2')\\
&&\times\iint_{\Lambda_1\times \Lambda_2}q_1(\lm_1)q_2(\lm_2)
P_A(a|x,\lm_1)P_B(b|y,\lm_1\lm_2)P_C(c|z,\lm_2){\rm{d}}\mu_1(\lm_1){\rm{d}}\mu_2(\lm_2)\\
&=&P(abc|xyz),\end{eqnarray*}
leading to Eq. (\ref{bi-p}). Similarly, Eq. (\ref{bi-p'}) follows from Eq.  (\ref{b22}). The proof is completed.

Moreover, when a CT ${\bf{P}}=\Lbrack P(abc|xyz)\Rbrack$ over $\Delta_3$ has a C-biLHVM (\ref{bi-MCT22}) given by counting measures $\mu_i$ on $\Lambda_i$ for all $i=1,2$, Eq. (\ref{bi-MCT22}) reduces to a discrete biLHVM (D-biLHVM):
\begin{equation}\label{Dbi-CT1}
P(abc|xyz)=\sum_{\lm_1=1}^{d_1}\sum_{\lm_2=1}^{d_2}q_1(\lm_1)q_2(\lm_2)
P_A(a|x,\lm_1)P_B(b|y,\lm_1\lm_2)P_C(c|z,\lm_2)
\end{equation}
where $q_i(\lm_i),P_A(a|x,\lm_1), P_B(b|y,\lm_1\lm_2)$ and $P_C(c|z,\lm_2)$  are PDs.

Conversely, if a CT ${\bf{P}}=\Lbrack P(abc|xyz)\Rbrack$ over $\Delta_3$ has a D-biLHVM (\ref{Dbi-CT1}), then it has a C-biLHVM (\ref{bi-MCT22}) given by counting measures $\mu_i$ on $\Lambda_i$ for all $i=1,2$. To show that the converse is also valid, we need to establish a decomposition lemma about a row-stochastic function matrix.

To give our results, some notations are necessary.
Put $N_X=(o_X)^{m_X}(X=A,B,C)$, which is the total number of maps from $[m_X]$ into $[o_X]$, and let  $$\Omega_A=\{J_1,J_2,\ldots,J_{N_A}\}=\{J|J:[m_A]\rightarrow[o_A]\},$$ $$\Omega_B=\{K_1,K_2,\ldots,K_{N_B}\}=\{K|K:[m_B]\rightarrow[o_B]\},$$ $$\Omega_C=\{L_1,L_2,\ldots,L_{N_C}\}=\{L|L:[m_C]\rightarrow[o_C]\}.$$
It is clear that every $m\times n$ $\{0,1\}$-row statistics matrix $T=[T_{ij}]$ corresponds uniquely to a mapping $F:[m]\rightarrow[n]$ so that $T_{ij}=\delta_{j,F(i)}$. Thus, the sets
of all  $\{0,1\}$- row-stochastic matrices of orders   $m_A\times o_A$,
 $m_B\times o_B$, and $m_C\times o_C$
can be written as
$$RSM^{(0,1)}_{m_A\times o_A}=\{R_i:=[\delta_{a,J_i(x)}]_{x,a}:i=1,2,\ldots,N_A\},$$
$$RSM^{(0,1)}_{m_B\times o_B}=\{K_j:=[\delta_{b,K_j(y)}]_{y,b}:j=1,2,\ldots,N_B\},$$
$$RSM^{(0,1)}_{m_C\times o_C}=\{L_k:=[\delta_{c,L_k(z)}]_{z,c}:k=1,2,\ldots,N_C\},$$
respectively.

{\bf Definition 2.2.} (1) An $m\times n$ real  matrix $B=[b_{ij}]$ is said to be row-stochastic (RS) if  $b_{ij}\ge0$ for all $i,j$ and $\sum_{j=1}^nb_{ij}(\lm)=1$  for all $i\in[m].$ Especially, $B=[b_{ij}]$ is said to be  $\{0,1\}$-RS if for each  row index $i$, there is a unique column index $J(i)$ such that $b_{ij}=\delta_{j,J(i)}$. We use $\{R_k:k\in[n^m]\}$ to denote the set of all  $m\times n$  $\{0,1\}$-RS matrices.

(2) An $m\times n$  function matrix $B(\lm)=[b_{ij}(\lm)]$ on a set $\Lambda$ is said to be RS if for each $\lm\in\Lambda$, $b_{ij}(\lm)\ge0$ for all $i,j$ and $\sum_{j=1}^nb_{ij}(\lm)=1$  for all $i\in[m].$

Recall that a {\it measurable space (MS)} \cite{Rudin} is a pair $(\Lambda,\Omega)$ of a set  $\Lambda$ and a $\sigma$-algebra $\Omega$ of subsets of $\Lambda$. When an MS $(\Lambda,\Omega)$ is given, the members of  $\Omega$ are called {\it $\Omega$-measurable sets} of $\Lambda$ and a function $f:\Lambda\rightarrow\R$ is said to be {\it $\Omega$-measurable} (measurable for short) if the inverse image $f^{-1}(G)$ of every open set $G$ in $\R$ is $\Omega$-measurable, i.e., an element of  $\Omega$. Furthermore, a {\it measure} $\mu$ on  $\Omega$ is a nonnegative extended real-valued function defined on the set  $\Omega$ satisfying the countable additivity:
$$A_i\in\Omega(i=1,2,\ldots),A_i\cap A_j=\O(i\ne j)\Rightarrow
\mu\left(\bigcup_{i=1}^\infty A_i\right)=\sum_{i=1}^\infty\mu(A_i).$$
Usually, we also assume that $\mu(A)<+\infty$ for some $A\in\Omega$.
A {\it measure space} \cite{Rudin} means a triple  $(\Lambda,\Omega,\mu)$ of a set  $\Lambda$, a $\sigma$-algebra $\Omega$ on $\Lambda$ and a measure $\mu$ on  $\Omega$.

{\bf Lemma 2.1.} {\it Let $(\Lambda,\Omega)$ be a measurable space and let  $m\times n$  function matrix $B(\lm)=[b_{ij}(\lm)]$ be  $\Omega$-measurable  on $\Lambda$, i.e., $b_{ij}$ are all $\Omega$-measurable functions on $\Lambda$. Then $B(\lm)$ is RS if and only if it can be written as:
\Bq{Lem3.2}
B(\lm)=\sum_{k=1}^{n^m}\alpha_k(\lm)R_k,\ \forall \lm\in\Lambda,\Eq
where $\alpha_k(k=1,2,\ldots,n^m)$ are all nonnegative and $\Omega$-measurable functions on $\Lambda$ with $\sum_{k=1}^{n^m}\alpha_k(\lm)=1$ for all $\lm\in\Lambda$.}

The sufficiency is clear and the necessity is proved in Appendix.

Based on this lemma, we have the following theorem, which gives a series of equivalent characterizations of bilocality of a CT. Especially, it says that
  a CT ${\bf{P}}$ over $\Delta_3$ has a C-biLHVM if and only if it has a D-biLHVM. Remarkably, characterization $(ii)$ in Theorem 2.1 was also given in
  \cite[Eq. (7)]{Branciard2012} by saying that ``it is well known  that Alice's local response function $P_A(a|x,\lm_1)$ can (without any loss of generality) be taken to be deterministic", based on the Fine's paper \cite{Fine}. Indeed, Fine in  \cite{Fine} did not give a mathematical proof of this conclusion, just a physical description.

{\bf Theorem 2.1.} {\it Let ${\bf{P}}=\Lbrack P(abc|xyz)\Rbrack$ be  a CT over $\Delta_3$. Then the following  statements $(i)$-$(v)$ are equivalent.

(i) ${\bf{P}}$ is bilocal, i.e., it has a C-biLHVM  (\ref{bi-MCT22}).

(ii) ${\bf{P}}$ has a D-biLHVM:
\begin{equation}\label{ddbi-T222}
P(abc|xyz)=\sum_{i=1}^{N_A}\sum_{j=1}^{N_B}\sum_{k=1}^{N_C}\pi({i,j,k}) \delta_{a,J_i(x)}\delta_{b,K_j(y)}\delta_{c,L_k(z)},\ \forall x,a,y,b,z,c,
\end{equation}
where
\Bq{2bi-PD}
\pi({i,j,k})=\iint_{\Lambda_1\times\Lambda_2}q_1(\lm_1)q_2(\lm_2)\alpha_i(\lm_1)\beta_j(\lm_1\lm_2)
\gamma_k(\lm_2){\rm{d}}\mu_1(\lm_1){\rm{d}}\mu_2(\lm_2),
\Eq
$q_1(\lm_1)$ and $q_2(\lm_2)$ are PDs of $\lm_1$ and $\lm_2$, respectively,  $\alpha_i(\lm_1)$, $\beta_j(\lm_1\lm_2)$ and $\gamma_k(\lm_2)$ are PDs of $i,j$ and $k$, respectively, and are measurable w.r.t. $\lm_1,(\lm_1\lm_2)$ and $\lm_2$, respectively.

(iii) ${\bf{P}}$ has a D-biLHVM:
\begin{equation}\label{Dbi-T222}
P(abc|xyz)=\sum_{i=1}^{N_A}\sum_{k=1}^{N_C}\pi_1(i)\pi_2(k) \delta_{a,J_i(x)}P_B(b|y,i,k)\delta_{c,L_k(z)},\ \forall x,a,y,b,z,c,
\end{equation}
where $q_i(\lm_i),P_A(a|x,\lm_1), P_B(b|y,\lm_1\lm_2), P_C(c|z,\lm_2)$ are PDs of $\lm_i,a,b,c,$ respectively.

(iv)  ${\bf{P}}$ is ``separable quantum", i.e., it can be generated by a family
\Bq{LMOVMs}
{\mathcal{M}}_{ABC}=\{M^{xyz}\}_{(x,y,z)\in[m_A]\times[m_B]\times[m_C]}
\Eq
of local POVMs $M^{xyz}=\{M_{a|x}\OO N_{b|y}\OO L_{c|z}\}_{(a,b,c)\in[o_A]\times[o_B]\times[o_C]}$ on a Hilbert space $\H_A\OO\H_B\OO\H_C$ together with a   pair $\{\rho_{AB_1},\rho_{B_2C}\}$ of {separable states} $\rho_{AB_1}$ and $\rho_{B_2C}$ of systems $\H_A\OO\H_{B_1}$ and $\H_{B_2}\OO\H_C$, respectively, in such a way that
\begin{equation}\label{Q-bilocal}
P(abc|xyz)=\tr[(M_{a|x}{\OO} N_{b|y}\OO L_{c|z})(\rho_{AB_1}\OO\rho_{B_2C})],\ \forall x,a,y,b,z,c.
\end{equation}

(v) ${\bf{P}}$ has   a D-biLHVM:
\begin{equation}\label{Dbi-CT}
P(abc|xyz)=\sum_{\lm_1=1}^{d_1}\sum_{\lm_2=1}^{d_2}q_1(\lm_1)q_2(\lm_2)
P_A(a|x,\lm_1)P_B(b|y,\lm_1\lm_2)P_C(c|z,\lm_2)
\end{equation}
where $q_i(\lm_i),P_A(a|x,\lm_1), P_B(b|y,\lm_1\lm_2), P_C(c|z,\lm_2)$ are PDs of $\lm_i,a,b,c,$ respectively.
}

{\bf Proof.} $(i)\Rightarrow(ii):$ Let $(i)$ be valid. Then the matrices
$$M_1(\lm_1):=[P_A(a|x,\lm_1)]_{x,a}, M_2(\lm_1\lm_2):=[P_B(b|y,\lm_1\lm_2)]_{y,b},
M_3(\lm_2)=:[P_C(c|z,\lm_2)]_{z,c}$$
are RS matrices for each parameter $\lm_k\in\Lambda_k$. It follows from Lemma 2.1 that they have the following decompositions:
$$M_1(\lm_1)=\sum_{i=1}^{N_A}\alpha_i(\lm_1)R_i,$$$$M_2(\lm_1\lm_2)
=\sum_{j=1}^{N_B}\beta_j(\lm_1\lm_2)Q_j,$$$$M_3(\lm_2)=\sum_{k=1}^{N_C}\gamma_k(\lm_2)S_k;$$
equivalently,
$$P(a|x,\lm)=\sum_{i=1}^{N_A}\alpha_i(\lm_1)\delta_{a,J_i(x)},$$$$
P(b|y,\lm_1\lm_2)=\sum_{j=1}^{N_B}\beta_j(\lm_1\lm_2)\delta_{b,K_j(y)}, $$$$
P(c|z,\lm_2)=\sum_{k=1}^{N_C}\gamma_k(\lm_2)\delta_{c,L_k(z)},
$$
where $\alpha_i(\lm_1)$, $\beta_j(\lm_1\lm_2)$ and $\gamma_k(\lm_2)$ are PDs of $i,j$ and $k$, respectively, and are measurable w.r.t. $\lm_1,(\lm_1\lm_2)$ and $\lm_2$, respectively.
Thus, Eq. (\ref{ddbi-T222}) follows from the C-biLHVM (\ref{bi-MCT22})
where $\pi({i,j,k})$ is given by (\ref{2bi-PD}). Thus, $(ii)$ is valid.

$(ii)\Rightarrow(iii):$
Let statement $(ii)$ be valid. For all $(i,k)\in[N_A]\times[N_C]$, put
$$
\pi_1(i)=\int_{\Lambda_1}q_1(\lm_1)\alpha_i(\lm_1){\rm{d}}\mu_1(\lm_1),
\pi_2(k)=\int_{\Lambda_2}q_2(\lm_2)\gamma_k(\lm_2){\rm{d}}\mu_2(\lm_2),$$
$$P_B(b|y,i,k)=\frac{1}{\pi_1(i)\pi_2(k)}
\iint_{\Lambda_1\times\Lambda_2}q_1(\lm_1)q_2(\lm_2)\alpha_i(\lm_1)
\gamma_k(\lm_2)P(b|y,\lm_1\lm_2){\rm{d}}\mu_1(\lm_1){\rm{d}}\mu_2(\lm_2)$$
for all $y\in[m_B],b\in[o_B]$ if $\pi_1(i)\pi_2(k)>0$; otherwise, define
$$P_B(b|y,i,k)=\frac{1}{o_B},\ \ \forall y\in[m_B], \forall b\in[o_B].$$
Then $\pi_1(i)$, $\pi_2(k)$ and $P_B(b|y,i,k)$ are PDs of $i,j$ and $b$, respectively, and
$$\iint_{\Lambda_1\times\Lambda_2}q_1(\lm_1)q_2(\lm_2)\alpha_i(\lm_1)
\gamma_k(\lm_2)P(b|y,\lm_1\lm_2){\rm{d}}\mu_1(\lm_1){\rm{d}}\mu_2(\lm_2)
=\pi_1(i)\pi_2(k)P_B(b|y,i,k)$$
for all $i,k,y,b$. Note that when $\pi_1(i)\pi_2(k)=0$, the left-hand side of above equation is less than or equal to $\pi_1(i)\pi_2(k)$ and must be $0$. It follows from Eqs. (\ref{ddbi-T222}) and (\ref{2bi-PD}) that $\forall x,a,y,b,z,c,$
\begin{eqnarray*}
P(abc|xyz)
&=&\sum_{i=1}^{N_A}\sum_{j=1}^{N_B}\sum_{k=1}^{N_C}\pi({i,j,k}) \delta_{a,J_i(x)}\delta_{b,K_j(y)}\delta_{c,L_k(z)}\\
&=&\sum_{i=1}^{N_A}\sum_{k=1}^{N_C}\delta_{a,J_i(x)}\delta_{c,L_k(z)}\\
&&\times\iint_{\Lambda_1\times\Lambda_2}q_1(\lm_1)q_2(\lm_2)\alpha_i(\lm_1)
\gamma_k(\lm_2)P(b|y,\lm_1\lm_2){\rm{d}}\mu_1(\lm_1){\rm{d}}\mu_2(\lm_2)\\
&=&\sum_{i=1}^{N_A}\sum_{k=1}^{N_C}\pi_1(i)\pi_2(k)\delta_{a,J_i(x)}P_B(b|y,i,k)\delta_{c,L_k(z)}.
\end{eqnarray*}
This shows that $(iii)$ is valid.

$(iii)\Rightarrow(iv):$ Let $(iii)$ be valid. By putting
$$\H_A=\H_{B_1}=\C^{N_A},\H_C=\H_{B_2}=\C^{N_C},\H_B=\H_{B_1}\OO\H_{B_2}=
\C^{N_A}\OO\C^{N_C},$$
taking orthonormal bases $\{|e_i\>\}_{i=1}^{N_A}$ and $\{|f_k\>\}_{k=1}^{N_C}$ for $\H_A$ and $\H_C$, respectively, defining POVMs:
$$M_{a|x}=\sum_{i=1}^{N_A}\delta_{a,J_i(x)}|e_i\>\<e_i|,L_{c|z}=\sum_{k=1}^{N_C}\delta_{c,L_k(z)}|f_k\>\<f_k|,$$
$$N_{b|y}=\sum_{i=1}^{N_A}\sum_{k=1}^{N_C}P_B(b|y,i,k)|e_i\>\<e_i|\OO|f_k\>\<f_k|,$$
and constructing separable states:
\Bq{RAB}\rho_{AB_1}=\sum_{s=1}^{N_A}\pi_1(s)
|e_s\>\<e_s|\OO|e_s\>\<e_s|,
\rho_{B_2C}=\sum_{t=1}^{N_C}\pi_2(t)|f_t\>\<f_t|\OO|f_t\>\<f_t|,\Eq
we get
$$\rho_{AB_1}\OO\rho_{B_2C}=\sum_{i=1}^{N_A}\sum_{k=1}^{N_C}
\pi_1(i)\pi_2(k)
|e_i\>\<e_i|\OO(|e_{i}\>\<e_{i}|\OO|f_k\>\<f_k|)\OO|f_{k}\>\<f_{k}|,$$
and then obtain Eq. (\ref{Q-bilocal}) from Eq. (\ref{Dbi-T222}).

$(iv)\Rightarrow(v):$ Let $(iv)$ be valid. Since  $\rho_{AB_1}$ and $\rho_{B_2C}$ are separable states of systems $AB_1$ and $B_2C$, they have   the following decompositions:
$$\rho_{AB_1}=\sum_{\lm_1=1}^{d_1}
q_1(\lm_1)|e_{\lm_1}\>\<e_{\lm_1}|\OO|f_{\lm_1}\>\<f_{\lm_1}|,$$
$$\rho_{B_2C}=\sum_{\lm_2=1}^{d_2}
q_2(\lm_2)|g_{\lm_2}\>\<g_{\lm_2}|\OO|h_{{\lm_2}}\>\<h_{{\lm_2}}|,$$
where   $q_i(\lm_i)$   is a PD of $\lm_i$,   $\{|e_{\lm_1}\>\}_{{\lm_1}=1}^{d_1}$,  $\{|f_{\lm_1}\>\}_{{\lm_1}=1}^{d_1}$, $\{|g_{\lm_2}\>\}_{\lm_2=1}^{d_2}$ and $\{|h_{{\lm_2}}\>\}_{\lm_2=1}^{d_2}$ are sets of pure states of $\H_A$,  $\H_{B_1}$, $\H_{B_2}$ and $\H_C$,  respectively. Thus,
$$\rho_{AB_1}\OO\rho_{B_2C}=\sum_{\lm_1=1}^{d_1}\sum_{\lm_2=1}^{d_2}
q_1(\lm_1)q_2(\lm_2)
|e_{\lm_1}\>\<e_{\lm_1}|\OO|f_{\lm_1}\>\<f_{\lm_1}|
\OO|g_{\lm_2}\>\<g_{\lm_2}|\OO|h_{\lm_2}\>\<h_{\lm_2}|
,$$
and so
 Eq. (\ref{Q-bilocal}) yields Eq. (\ref{Dbi-CT}) by taking
$P_B(b|y,\lm_1\lm_2)=\<f_{\lm_1}g_{\lm_2}|N_{b|y}|f_{\lm_1}g_{\lm_2}\>,$ and
$$P_A(a|x,\lm_1)=\<e_{\lm_1}|M_{a|x}|e_{\lm_1}\>,
P_C(c|z,\lm_2)=\<h_{\lm_2}|L_{c|z}|h_{\lm_2}\>.
$$

$(v)\Rightarrow(i):$ When Eq. (\ref{Dbi-CT}) holds, Eq. (\ref{bi-MCT22}) is valid for $\Lambda_i=[d_i]$, $\Omega_i=2^{\Lambda_i}$ (power set of $\Lambda_i$) and the counting measures $\mu_i$ on $\Lambda_i(i=1,2)$.  Hence, ${\bf{P}}$ is bilocal. The proof is completed.

{\bf Proposition 2.2.} {\it A CT ${\bf{P}}=\Lbrack P(abc|xyz)\Rbrack$ over $\Delta_3$ is bilocal if and only if it can be written as
\begin{equation}\label{P2.2}
P(abc|xyz)=\sum_{i=1}^{N_A}\sum_{j=1}^{N_B}\sum_{k=1}^{N_C}\pi_1(i)\pi_2(k)
p(j|i,k)\delta_{a,J_i(x)}\delta_{b,K_j(y)}\delta_{c,L_k(z)}
\end{equation}
for all $x,y,z,a,b,c,$
where $\pi_1(i),\pi_2(k)$ and $p(j|i,k)$ are PDs of $i,k$ and $j$, respectively.}

{\bf Proof.} When (\ref{P2.2}) is valid, letting
\Bq{P221}
P(b|y,i,k)=\sum_{j=1}^{N_B}p(j|i,k)\delta_{b,K_j(y)}
\Eq
yields Eq. (\ref{Dbi-T222}). It follows from Theorem 2.1 that ${\bf{P}}\in\mathcal{CT}^{\textrm{bilocal}}(\Delta_3)$. Conversely, when ${\bf{P}}\in\mathcal{CT}^{\textrm{bilocal}}(\Delta_3)$, Theorem 2.1 yields that Eq. (\ref{Dbi-T222}) holds. By considering measurable space $([N_A]\times[N_C], P([N_A]\times[N_C]))$ and then using Lemma 2.1 for the matrix $M(i,k)=[P(b|y,i,k)]$ with $(y,b)$-entry $P(b|y,i,k)$, we conclude the composition (\ref{P221}) of  $P(b|y,i,k)$ in (\ref{Dbi-T222}). Thus, (\ref{Dbi-T222}) becomes (\ref{P2.2}). The proof is completed.

Note that \cite[Theorem 5.1]{BaiLH} every ${\bf{P}}=\Lbrack P(abc|xyz)\Rbrack\in\mathcal{CT}^{\textrm{Bell-local}}(\Delta_3)$ if and only if it can be written as
\begin{equation}\label{b222}
P(abc|xyz)=\sum_{i=1}^{N_A}\sum_{j=1}^{N_B}\sum_{k=1}^{N_C}q_{ijk} \delta_{a,J_i(x)}\delta_{b,K_j(y)}\delta_{c,L_k(z)},\ \forall x,a,y,b,z,c,
\end{equation} where $\{q_{ijk}\}$ is a PD of $i,j,k$.
Thus, from the decomposition (\ref{P2.2}) of a bilocal CT, we see that every bilocal CT is Bell local, but not the inverse.
Based on decomposition (\ref{b222}), it is easy to check that  the set $\mathcal{CT}^{\textrm{Bell-local}}(\Delta_3)$  is a convex compact set, and so
$$\mathcal{CT}^{\textrm{Bell-local}}(\Delta_3)\supset{\rm{conv}}\left(\mathcal{CT}^{\textrm{bilocal}}(\Delta_3)\right).$$
 Remark 2.1 yields that ${\bf{D}}_{ijk}:=\Lbrack \delta_{a,J_i(x)}\delta_{b,K_j(y)}\delta_{c,L_k(z)}\Rbrack\in \mathcal{CT}^{\textrm{bilocal}}(\Delta_3)$  for all $i,j,k.$ Thus, Eq.  (\ref{b222}) implies that
\Bq{BLbiL}\mathcal{CT}^{\textrm{Bell-local}}(\Delta_3)={\rm{conv}}\left(\mathcal{CT}^{\textrm{bilocal}}(\Delta_3)\right),\Eq
which was pointed out in \cite[II.B]{Branciard2012}.

Since the hidden variables $\lm_1$ and $\lm_2$ in a C-biLHVM (\ref{bi-MCT22}) or a D-triLHVM (\ref{Dbi-CT}) are assumed to be independent, the sets  $\mathcal{CT}^{\textrm{bilocal}}(\Delta_3)$ is not necessarily convex (Remark 2.2). Branciard et al. claimed in \cite[Appendix A.1]{Branciard2012} that
 the bilocal set is connected. Next, we give a detail proof of the last  conclusion.

\begin{figure}[h]
  \centering
  \includegraphics[width=10cm]{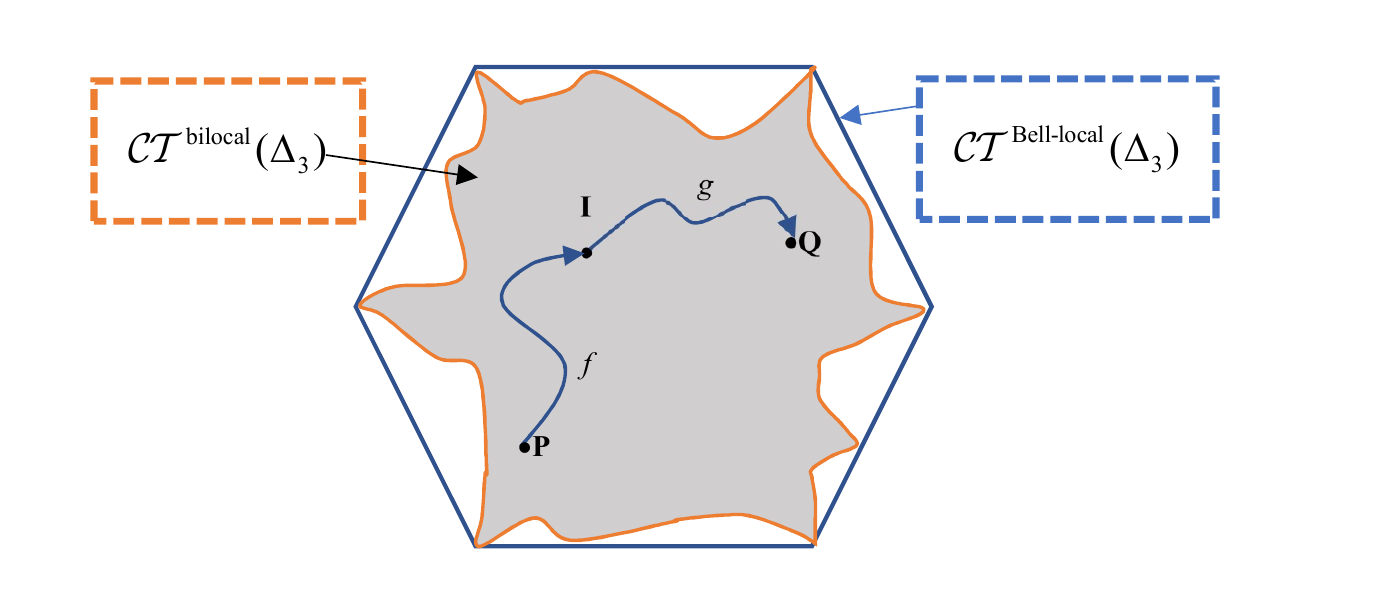}\\
  \caption{Path-connectness of the set $\mathcal{CT}^{\textrm{bilocal}}(\Delta_3)$}\label{3-path-connectness}
\end{figure}

{\bf Corollary 2.1.}{(Path-connectedness)}\cite[Appendix A]{Branciard2012}  {\it The set}\ $\mathcal{CT}^{\textrm{bilocal}}(\Delta_3)$ {\it is  path-connected  in the Hilbert space $\mathcal{T}(\Delta_3)$. See Fig. \ref{3-path-connectness}.}

{\bf Proof.} Let ${\bf{P}}=\Lbrack P(abc|xyz)\Rbrack$ and ${\bf{Q}}=\Lbrack Q(abc|xyz)\Rbrack$ be any two elements of $\mathcal{CT}^{\textrm{bilocal}}(\Delta_3)$. Then Theorem 2.1 implies that  ${\bf{P}}$ and ${\bf{Q}}$  have  D-biLHVMs:
\begin{equation}\label{1Dbi-CT}
P(abc|xyz)=\sum_{\lm_1=1}^{d_1}\sum_{\lm_2=1}^{d_2}p_1(\lm_1)p_2(\lm_2)
P_A(a|x,\lm_1)P_B(b|y,\lm_1\lm_2)P_C(c|z,\lm_2)
\end{equation}
where
 $p_i(\lm_i), P_A(a|x,\lm_1),$ $P_B(b|y,\lm_1\lm_2),P_C(c|z,\lm_2)$ are PDs of $\lm_i,a,b,c,$ respectively.,
and
\begin{equation}\label{1Dbi-CT}
Q(abc|xyz)=\sum_{\xi_1=1}^{d_1'}\sum_{\xi_2=1}^{d_2'}q_1(\xi_1)q_2(\xi_2)
Q_A(a|x,\xi_1)Q_B(b|y,\xi_1\xi_2)Q_C(c|z,\xi_2)
\end{equation}
where $q_i(\xi_i), Q_A(a|x,\xi_1),  Q_B(b|y,\xi_1\xi_2),Q_C(c|z,\xi_2)$ are PDs of $\xi_i,a,b,c,$ respectively.

Put $I(abc|xyz)\equiv\frac{1}{o_Ao_Bo_C}$, then ${\bf{I}}:=\Lbrack I(abc|xyz)\Rbrack\in\mathcal{CT}^{\textrm{bilocal}}(\Delta_3).$ For every $t\in[0,1/2]$, set
$$P^t_1(a|x,\lm_1)=(1-2t)P_1(a|x,\lm_1)+2t\frac{1}{o_A};$$
$$P^t_2(b|y,\lm_1\lm_2)=(1-2t)P_2(b|y,\lm_1\lm_2)+2t\frac{1}{o_B};$$
$$P^t_3(c|z,\lm_2)=(1-2t)P_3(c|z,\lm_2)+2t\frac{1}{o_C},$$
which are clearly PDs of $a,b$ and $c$, respectively. Putting
\begin{eqnarray*}
P^t(abc|xyz)=\sum_{\lm_1=1}^{d_1}\sum_{\lm_2=1}^{d_2}p_1(\lm_1)p_2(\lm_2)
P^t_1(a|x,\lm_1)P^t_2(b|y,\lm_1\lm_2)P^t_3(c|z,\lm_2)
\end{eqnarray*}
yields  a bilocal CT
${f}(t):=\Lbrack P^t(abc|xyz)\Rbrack$ over $\Delta_3$ for all $t\in[0,1/2]$ with ${f}(0)={\bf{P}}$ and ${f}(1/2)={\bf{I}}$. Obviously, the map $t\mapsto {f}(t)$ from $[0,1/2]$ into $\mathcal{CT}^{\textrm{bilocal}}(\Delta_3)$ is continuous.

Similarly, for every $t\in[1/2,1]$, set
$$Q^t_1(a|x,\xi_1)=(2t-1)Q_1(a|x,\xi_1)+2(1-t)\frac{1}{o_A};$$
$$Q^t_2(b|y,\xi_1\xi_2)=(2t-1)Q_2(b|y,\xi_1\xi_2)+2(1-t)\frac{1}{o_B};$$
$$Q^t_3(c|z,\xi_2)=(2t-1)Q_3(c|z,\xi_2)+2(1-t)\frac{1}{o_C},$$
which are clearly PDs of $a,b$ and $c$, respectively. Put
\begin{eqnarray*}
Q^t(abc|xyz)=\sum_{\xi_1=1}^{d_1'}\sum_{\xi_2=1}^{d_2'}q_1(\xi_1)q_2(\xi_2)
Q^t_1(a|x,\xi_1)Q^t_2(b|y,\xi_1\xi_2)Q^t_3(c|z,\xi_2),
\end{eqnarray*}
then
${g}(t):=\Lbrack Q^t(abc|xyz)\Rbrack$ is a bilocal CT over $\Delta_3$ for all $t\in[1/2,1]$ with ${g}(1/2)={\bf{I}}$ and ${g}(1)={\bf{Q}}$. Obviously, the map $t\mapsto {g}(t)$ from $[1/2,1]$ into $\mathcal{CT}^{\textrm{bilocal}}(\Delta_3)$ is continuous. Thus, the function
$f:[0,1]\rightarrow \mathcal{CT}^{\textrm{bilocal}}(\Delta_3)$ defined by
$$p(t)=\left\{\begin{array}{cc}
                {f}(t), & t\in[0,1/2];\\
                {g}(t), & t\in(1/2,1],
              \end{array}\right.$$
is continuous everywhere
 and then induces a path $p$ in $\mathcal{CT}^{\textrm{bilocal}}(\Delta_3)$, connecting ${\bf{P}}$ and ${\bf{Q}}$. This shows that $\mathcal{CT}^{\textrm{bilocal}}(\Delta_3)$ is path-connected.  The proof is completed.

Next, let us discuss the star-convexity of the set of all bilocal CTs over $\Delta_3$ such that Alice has a fixed distribution. Recall that a subset $D$ of a vector space $V$ is said to star-convex if it has an element $s$, called a sun of $S$, such that $(1-t)s+tD\subset D$ for all $t\in[0,1].$

 To do this,  take a CT ${\bf{E}}=\Lbrack E(a|x)\Rbrack$ over $[o_A]\times[m_A]$, we let
$$\mathcal{CT}_{A-{\bf{E}}}^{\textrm{bilocal}}(\Delta_3)=
\left\{{\bf{P}}\in \mathcal{CT}^{\textrm{bilocal}}(\Delta_3):\ {\bf{P}}_A={\bf{E}}\right\}.$$

The proof of the following corollary is suggested by an argument in \cite[Appendix A.2]{Branciard2012}.

\begin{figure}[h]
  \centering
  \includegraphics[width=10cm]{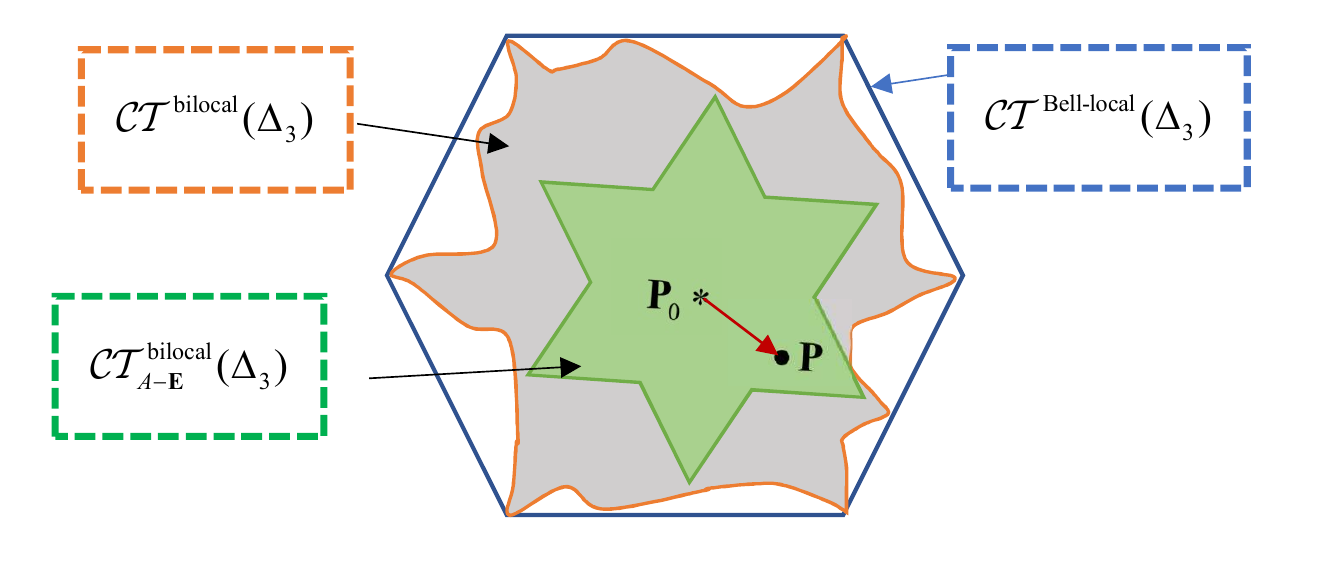}\\
  \caption{Star-convexity of the set $\mathcal{CT}_{A-{\bf{E}}}^{\textrm{bilocal}}(\Delta_3)$.}\label{3-star-convex}
\end{figure}

{\bf Corollary 2.2.}(Star-convexity)\cite[Appendix A.2]{Branciard2012} {\it The set} $\mathcal{CT}_{A-{\bf{E}}}^{\textrm{bilocal}}(\Delta_3)$ {\it is star-convex the Hilbert space $\mathcal{T}(\Delta_3)$. See Fig. \ref{3-star-convex}.}

{\bf Proof.} Put $P_0(abc|xyz)=E(a|x)\times\frac{1}{o_B}\times\frac{1}{o_C}$, then ${\bf{P}}_0:=\Lbrack P_0(abc|xyz)\Rbrack$ is a bilocal  CT over $\Delta_3$ (Remark 2.1). Next we show that ${\bf{P}}_0$ is a sun of $\mathcal{CT}_{A-{\bf{E}}}^{\textrm{bilocal}}(\Delta_3)$, i.e.,
\Bq{ST}(1-t){\bf{P}}_0+t\mathcal{CT}_{A-{\bf{E}}}^{\textrm{bilocal}}(\Delta_3)\subset \mathcal{CT}_{A-{\bf{E}}}^{\textrm{bilocal}}(\Delta_3),\ \forall t\in[0,1].\Eq
Let
 ${\bf{P}}=\Lbrack P(abc|xyz)\Rbrack\in \mathcal{CT}_{A-{\bf{E}}}^{\textrm{bilocal}}(\Delta_3)$. We see from  Theorem 2.1 that  ${\bf{P}}$ has a  D-biLHVM:
\begin{equation}\label{1Dbi-CT}
P(abc|xyz)=\sum_{\lm_1=1}^{d_1}\sum_{\lm_2=1}^{d_2}p_1(\lm_1)p_2(\lm_2)
P_A(a|x,\lm_1)P_B(b|y,\lm_1\lm_2)P_C(c|z,\lm_2).
\end{equation}
The condition $P_A(a|x)=E(a|x)$ becomes
\Bq{PaP}
\sum_{\lm_1=1}^{d_1}p_1(\lm_1)P_A(a|x,\lm_1)=E(a|x),\ \forall x\in[m_A],a\in[o_A].
\Eq
For every $t\in[0,1]$, set
$$
f(\lm_2,s)=\left\{\begin{array}{cc}
p_2(\lm_2)(1-t), & s=0; \\
                    p_2(\lm_2)t, & s=1,
                  \end{array}\right.
$$$$
P_B(b|y,\lm_1,(\lm_2,s))=\left\{\begin{array}{cc}
                    \frac{1}{o_B}, & s=0; \\
                    P_B(b|y,\lm_1\lm_2), & s=1,
                  \end{array}\right.
$$$$
P_C(c|z,(\lm_2,s))=\left\{\begin{array}{cc}
                    \frac{1}{o_C}, & s=0; \\
                    P_C(c|z,\lm_2), & s=1,
                  \end{array}\right.
$$
which are PDs of $(\lm_2,s),b$ and $c$, respectively.
For all $x,y,z,a,b,z$, we see from (\ref{PaP}) that
\begin{eqnarray*}
&&\sum_{\lm_1\in[d_1]}\sum_{(\lm_2,s)\in[d_2]\times\{0,1\}}p_1(\lm_1)f(\lm_2,s)
P_A(a|x,\lm_1)\\
&&\times P_B(b|y,\lm_1(\lm_2,s))P_C(c|z,\lm_2(\lm_2,s))\\
&=&\sum_{\lm_1\in[d_1]}\sum_{\lm_2\in[d_2]}p_1(\lm_1)p_2(\lm_2)(1-t)
P_A(a|x,\lm_1)\frac{1}{o_B}\frac{1}{o_C}\\
&&+\sum_{\lm_1\in[d_1]}\sum_{\lm_2\in[d_2]}p_1(\lm_1)p_2(\lm_2)t
P_A(a|x,\lm_1)P_B(b|y,\lm_1\lm_2)P_C(c|z,\lm_2)\\
&=&(1-t)P_0(abc|xyz)+tP(abc|xyz).
\end{eqnarray*}
This shows that ${\bf{Q}}:=(1-t){\bf{P}}_0+t{\bf{P}}$ is bilocal with ${\bf{Q}}_A={\bf{E}}$ and then an element of  ${\bf{P}}\in\mathcal{CT}_{A-{\bf{E}}}^{\textrm{bilocal}}(\Delta_3)$.
This shows that (\ref{ST}) holds and so $\mathcal{CT}_{A-{\bf{E}}}^{\textrm{bilocal}}(\Delta_3)$ is star-convex with a sun ${\bf{P}}_0$.  The proof is completed.

Similarly, for any fixed CT ${\bf{F}}=\Lbrack F(c|z)\Rbrack$ over $[o_C]\times[m_C]$, the set
$$\mathcal{CT}_{C-{\bf{F}}}^{\textrm{bilocal}}(\Delta_3)=
\left\{{\bf{P}}\in \mathcal{CT}^{\textrm{bilocal}}(\Delta_3):\ {\bf{P}}_C={\bf{F}}\right\}$$
is also star-convex.

Another application of Theorem 2.1 is to prove the compactness of $\mathcal{CT}^{\textrm{bilocal}}(\Delta_3)$.

{\bf{Corollary 2.3.}}(Compactness) {\it The set}  $\mathcal{CT}^{\textrm{bilocal}}(\Delta_3)$ {\it is  compact in the Hilbert space $\mathcal{T}(\Delta_3)$.}

{\bf{Proof.}} Let $\textbf{P}\in\mathcal{CT}(\Delta_3)$, $\{{\bf{P}}_n\}_{n=1}^\infty\subset\mathcal{CT}^{\textrm{bilocal}}(\Delta_3)$ with $\textbf{P}_n\rightarrow \textbf{P}$ as $n\rightarrow\infty$, i.e., $P_n(abc|xyz)\rightarrow P(abc|xyz)$  as $n\rightarrow\infty$ for all $(a,b,c,x,y,z)\in\Delta_3$. According to Theorem 2.1, we may assume that
\begin{equation}\label{ndDbi-T222}
P_n(abc|xyz)=\sum_{i=1}^{N_A}\sum_{k=1}^{N_C}\pi^{(n)}_1(i)\pi^{(n)}_2(k) \delta_{a,J_i(x)}P^{(n)}_B(b|y,i,k)\delta_{c,L_k(z)}
\end{equation}
for all $x,a,y,b,z,c,$ where, for each $n=1,2,\ldots,$
$$\{\pi^{(n)}_1(i)\}_{i\in[N_A]},\ \{\pi^{(n)}_2(k)\}_{k\in[N_C]}, {\rm{\ and\ }} \{P^{(n)}_B(b|y,i,k)\}_{b\in[o_B]}(\forall y,i,k)$$ are PDs of $i,k,b$, respectively.
By taking subsequences if necessary, we may assume that
$$\lim_{n\rightarrow\infty}\pi^{(n)}_1(i)=\pi_1(i)(\forall i),\ \lim_{n\rightarrow\infty}\pi^{(n)}_2(k)=\pi_2(k)(\forall k),$$ and
$$\lim_{n\rightarrow\infty}P^{(n)}_B(b|y,i,k)=P_B(b|y,i,k)(\forall y,b,i,k).$$
Obviously, $\{\pi_1(i)\}_{i\in[N_A]}$ and $\{\pi_2(k)\}_{k\in[N_C]}$, and $\{P_B(b|y,i,k)\}_{b\in[o_B]}(\forall y,i,k)$ are PDs of $i,k,b$. Letting $n\rightarrow\infty$ in Eq. (\ref{ndDbi-T222})  yields that
$$
P(abc|xyz)=\sum_{i=1}^{N_A}\sum_{k=1}^{N_C}\pi_1(i)\pi_2(k) \delta_{a,J_i(x)}P_B(b|y,i,k)\delta_{c,L_k(z)},\ \forall x,a,y,b,z,c.
$$
 Using Theorem 2.1 again implies that
${\bf{P}}\in\mathcal{CT}^{\textrm{bilocal}}(\Delta_3)$. Since $\mathcal{CT}^{\textrm{bilocal}}(\Delta_3)$ is also bounded, it is compact. The proof is completed.

Similar to Definition 2.1  we can introduce and discuss bilocality of a tripartite CT ${\bf{P}}=\Lbrack P(abc|xyz)\Rbrack$  over $\Delta_3$ defined by \begin{eqnarray}\label{A-bi-CT22}
P(abc|xyz)&=&\iint_{\Lambda_1\times\Lambda_2}q_1(\lm_1)q_2(\lm_2)
P_A(a|x,\lm_1\lm_2)\nonumber\\&&\times P_B(b|y,\lm_1)P_C(c|z,\lm_2){\rm{d}}\mu_1(\lm_1){\rm{d}}\mu_2(\lm_2)
\end{eqnarray}
or
\begin{eqnarray}\label{C-bi-CT22}
P(abc|xyz)&=&\iint_{\Lambda_1\times\Lambda_2}q_1(\lm_1)q_2(\lm_2)
P_A(a|x,\lm_1)\nonumber\\
&&\times P_B(b|y,\lm_2)P_C(c|z,\lm_1\lm_2){\rm{d}}\mu_1(\lm_1){\rm{d}}\mu_2(\lm_2).
\end{eqnarray}

The last type here of bilocality is just the special case of $n$-locality for  $n=2$, i.e., the $2$-local case, but not the first one.

\section{Bilocality of probability tensors}
\setcounter{equation}{0}

The usual Bell nonlocality of a quantum state or a quantum network is the property that is exhibited by performing a set of non-compatible local POVM measurements. Renou et al. \cite{Renou}  pointed out that
quantum nonlocality in a quantum network can be demonstrated  without the need of having various input settings, but only by considering the joint statistics  of fixed local measurement outputs. For example,  for a tripartite network, it suffices to consider the tripartite probability distribution $\{P(a,b,c)\}$ of local measurement outcomes $a,b,c$, where $a\in[o_A],b\in[o_B]$ and $c\in[o_C]$. These probabilities form a tensor ${\bf{P}}=\Lbrack P(a,b,c)\Rbrack$  over $O_3=[o_A]\times[o_B]\times[o_C]$, we call it a {\it probability tensor} (PT)\cite{Fritz2012}. Note that  every PT ${\bf{P}}=\Lbrack P(a,b,c)\Rbrack$  over $O_3$ can be written as
$$P(a,b,c)=\sum_{\lm=(\lm_1,\lm_2,\lm_3)\in O_3}q(\lm)P_A(a|\lm)P_B(b|\lm)P_C(c|\lm) (\forall a,b,c),$$
where
$$q(\lm)=P(\lm_1,\lm_2,\lm_3),P_A(a|\lm)=\delta_{a,\lm_1},
P_B(b|\lm)=
\delta_{b,\lm_2},P_C(c|\lm)=\delta_{c,\lm_3},$$
which a PDs of $\lm,a,b$ and $c$. Thus, every PT ${\bf{P}}$ is always Bell local.

{\bf Definition 3.1.} A PT  ${\bf{P}}=\Lbrack P(a,b,c)\Rbrack$  over $O_3$ is said to {\it bilocal} if it admits a C-biLHVM:
{\begin{equation}\label{bi-PT22}
P(a,b,c)=\iint_{\Lambda_1\times\Lambda_2}q_1(\lm_1)q_2(\lm_2)
P_A(a|\lm_1)P_B(b|\lm_1\lm_2)P_C(c|\lm_2){\rm{d}}\mu_1(\lm_1){\rm{d}}\mu_2(\lm_2)
\end{equation}}
for a measure space $(\Lambda_1\times\Lambda_2,\Omega_1\times\Omega_2,\mu_1\times\mu_2)$, where

(a) $q_1(\lm_1)$ and $P_A(a|\lm_1)(a\in[o_A])$ are nonnegative measurable on $\Lambda_1$,  $q_2(\lm_2)$ and $P_C(c|\lm_2)(c\in[o_C])$ are nonnegative measurable on $\Lambda_2$, and $P_B(b|\lm_1\lm_2)(b\in[o_B])$ are nonnegative measurable on $\Lambda_1\times\Lambda_2$;

(b) $q_i(\lm_i),P_A(a|\lm_1), P_B(b|\lm_1\lm_2)$ and $P_C(c|\lm_2)$ are  PDs  of $\lm_i,a,b,c,$ respectively.

We call Eq. (\ref{bi-PT22}) a  C-biLHVM  of ${\bf{P}}$ and denote by  $\mathcal{PT}^{\textrm{bilocal}}(O_3)$ the set of all bilocal PTs over $O_3$.

For example, the PT ${\bf{p}}=\Lbrack\pi({i,j,k})\Rbrack$ given by Eq. (\ref{2bi-PD}) is bilocal.

{\bf Remark 3.1.} From Definition 3.1, we observe that when  a PT ${\bf{P}}=\Lbrack P(a,b,c)\Rbrack$  over $O_3$ is  bilocal, its marginal distributions satisfy:
\Bq{pNec}P_{AC}(a,c)=P_{A}(a)P_{C}(c),\ \ \forall a,c.\Eq
By using this property of a bilocal PT, we can find that not all PTs over   $O_3$ are  bilocal.

{\bf Example 3.1.} Let $o_X=2(X=A,B,C)$ and define a PT ${\bf{P}}=\Lbrack P(a,b,c)\Rbrack$ by
$$P(1,1,1)=P(2,2,2)=\frac{1}{2}, P(a,b,c)=0 {\rm{\ if \ }} (a,b,c)\ne(1,1,1) {\rm{\ or\ }} (a,b,c)\ne(2,2,2).$$
Then
$$P_{AC}(1,1)=P_{AC}(2,2)=1/2,P_{AC}(a,c)=0 {\rm{\ if \ }} (a,c)\ne(1,1) {\rm{\ or\ }} (a,c)\ne(2,2),$$
$$P_{A}(1)=P_{A}(2)=1/2,P_{C}(1)=P_{C}(2)=1/2,$$
while
$$P_{AC}(1,1)=\frac{1}{2}\ne\frac{1}{4}=P_{A}(1)P_{C}(1).$$
Thus, ${\bf{P}}\notin\mathcal{PT}^{\textrm{bilocal}}(O_3)$.

 Clearly, a PT ${\bf{P}}=\Lbrack P(a,b,c)\Rbrack$  over $O_3$ is bilocal if and only if the CT
  ${\bf{P}}_1:=\Lbrack P(abc|111)\Rbrack$  over $\Delta_3=[o_A]\times[o_B]\times[o_C]\times[1]\times[1]\times[1]$ is C-bilocal, in this case $N_A=o_A,N_B=o_B,N_C=o_C$. Thus, as a special case of Theorem 2.1, we have the following.

 {\bf Theorem 3.1.} {\it For a PT ${\bf{P}}=\Lbrack P(a,b,c)\Rbrack$  over $O_3$, the following  statements $(i)$-$(iv)$ are equivalent.

(i) ${\bf{P}}$ is bilocal, i.e., it has a C-biLHVM (\ref{bi-PT22}).

(ii) ${\bf{P}}$ has a D-biLHVM:
\begin{equation}\label{pDbi-T222}
P(a,b,c)=\sum_{i=1}^{o_A}\sum_{k=1}^{o_C}\pi_1(i)\pi_2(k) \delta_{a,i}P_B(b|i,k)\delta_{c,k},\ \forall a,b,c,
\end{equation}
where $\pi_1(i)$, $\pi_2(k)$, and  $P_B(b|i,k)$ are PDs of $i,k,$ and $b$, respectively.

(iii)  ${\bf{P}}$ is ``separable quantum", i.e., it can be generated by a local POVM
 $M=\{M_{a}\OO N_{b}\OO L_{c}\}_{(a,b,c)\in[o_A]\times[o_B]\times[o_C]}$ on the Hilbert space $\H_A\otimes(\H_{B_1}\OO\H_{B_2})\OO\H_C$ together with a pair $\{\rho_{AB_1},\rho_{B_2C}\}$ of {separable states} $\rho_{AB_1}$ and $\rho_{B_2C}$ of systems $\H_A\otimes\H_{B_1}$ and $\H_{B_2}\OO\H_C$, respectively, in such a way that
\begin{equation}\label{pQ-bilocal}
P(a,b,c)=\tr[(M_{a}\OO N_{b}\OO L_{c})(\rho_{AB_1}\OO\rho_{B_2C})],\ \forall a,b,c.
\end{equation}

(iv) ${\bf{P}}$ has a D-biLHVM:
\begin{equation}\label{Pbi-CT}
P(a,b,c)=\sum_{\lm_1=1}^{d_1}\sum_{\lm_2=1}^{d_2}q_1(\lm_1)q_2(\lm_2)
P_A(a|\lm_1)P_B(b|\lm_1\lm_2)P_C(c|\lm_2)
\end{equation}
where $q_i(\lm_i), P_A(a|\lm_1)$,  $P_B(b|\lm_1\lm_2)$, and $P_C(c|\lm_2)$ are PDs of $\lm_i,a,b,$ and $c$, respectively.
}

Motivated by \cite{Renou}, we call a PT  ${\bf{P}}=\Lbrack P(a,b,c)\Rbrack$  over $O_3$ to be {\it trilocal} if it can be written as
\begin{eqnarray}\label{P23}
 P(a,b,c)=
\sum_{\lm_1=1}^{n_1}\sum_{\lm_2=1}^{n_2}\sum_{\lm_3=1}^{n_3}q_1(\lm_1)q_2(\lm_2)q_3(\lm_3)
P_A(a|\lm_3\lm_1)P_B(b|\lm_1\lm_2)P_C(c|\lm_2\lm_3)
\end{eqnarray}
for all $a\in[o_A],b\in[o_B],c\in[o_C]$, where
$$\{q_k(\lm_k)\}_{\lm_k\in[n_k]},
\{P_A(a|\lm_3\lm_1)\}_{a\in[o_A]},
\{P_B(b|\lm_1\lm_2)\}_{b\in[o_B]},\{P_C(c|\lm_2\lm_3)\}_{c\in[o_C]}$$ are PDs.
Such a PT $P$ is also called to be {\it classical in the $C_3$ scenario},  according to \cite[Definition 2.12]{Fritz2012}. Let ${\bf{P}}=\Lbrack P(a,b,c)\Rbrack$  over $O_3$ be a bilocal PT. Then it has a D-biLHVM (\ref{Pbi-CT}), which implies  (\ref{P23}) by putting  $$\lm_3=1,q_3(\lm_3)=1,P_A(a|\lm_3\lm_1)=P_A(a|\lm_1),P_C(c|\lm_2\lm_3)=P_C(c|\lm_2).$$
Hence, it is trilocal.  It was proved in \cite{Renou} that there exists a quantum nontrilocal PT ${\bf{P}}=\Lbrack P(a,b,c)\Rbrack$  over $O_3=[2]\times[4]\times[2]$. Thus, there exists a quantum PT that is non bilocal.

As special cases of Corollaries 2.1, 2.2 and 2.3, we have the following  conclusions.

{\bf{Corollary 3.1.}} {\it The set $\mathcal{PT}^{\textrm{bilocal}}(O_3)$  is  path-connected w.r.t. the pointwise-convergence topology on  $O_3$.}

By taking PDs ${\bf{E}}=\{E(a)\}_{a\in[o_A]}$ and ${\bf{F}}=\{F(c)\}_{c\in[o_C]}$, letting
$$\mathcal{PT}_{A-{\bf{E}}}^{\textrm{bilocal}}(O_3)=
\left\{{\bf{P}}\in \mathcal{PT}^{\textrm{bilocal}}(O_3):\ {\bf{P}}_A={\bf{E}}\right\},$$
$$\mathcal{PT}_{C-{\bf{F}}}^{\textrm{bilocal}}(O_3)=
\left\{{\bf{P}}\in \mathcal{PT}^{\textrm{bilocal}}(O_3):\ {\bf{P}}_C={\bf{F}}\right\},$$ we obtain have the following.

{\bf{Corollary 3.2.}} {\it The sets $\mathcal{PT}_{A-{\bf{E}}}^{\textrm{bilocal}}(O_3)$ and $\mathcal{PT}_{C-{\bf{F}}}^{\textrm{bilocal}}(O_3)$ and are star-convex.}

{\bf{Corollary 3.3.}} {\it The set $\mathcal{PT}^{\textrm{bilocal}}(O_3)$  is a compact subset of the set $\mathcal{PT}(O_3)$ of all PTs over $O_3$ w.r.t. the pointwise-convergence topology on  $O_3$.}

Similar to Definition 3.1  we can introduce and discuss bilocality of a tripartite PT
 ${\bf{P}}=\Lbrack P(a,b,c)\Rbrack$  over $O_3$ described by
\begin{equation}\label{A-bi-PT22}
P(a,b,c)=\iint_{\Lambda_1\times\Lambda_2}q_1(\lm_1)q_2(\lm_2)
P_A(a|\lm_1\lm_2)P_B(b|\lm_1)P_C(c|\lm_2){\rm{d}}\mu_1(\lm_1){\rm{d}}\mu_2(\lm_2),
\end{equation}
or
\begin{equation}\label{C-bi-PT22}
P(a,b,c)=\iint_{\Lambda_1\times\Lambda_2}q_1(\lm_1)q_2(\lm_2)
P_A(a|\lm_1)P_B(b|\lm_2)P_C(c|\lm_1\lm_2){\rm{d}}\mu_1(\lm_1){\rm{d}}\mu_2(\lm_2).
\end{equation}

\section{$n$-Locality of $n+1$-partite CTs}
\setcounter{equation}{0}\setcounter{section}{4}

\begin{figure}[h]
  \centering
  \includegraphics[width=7cm]{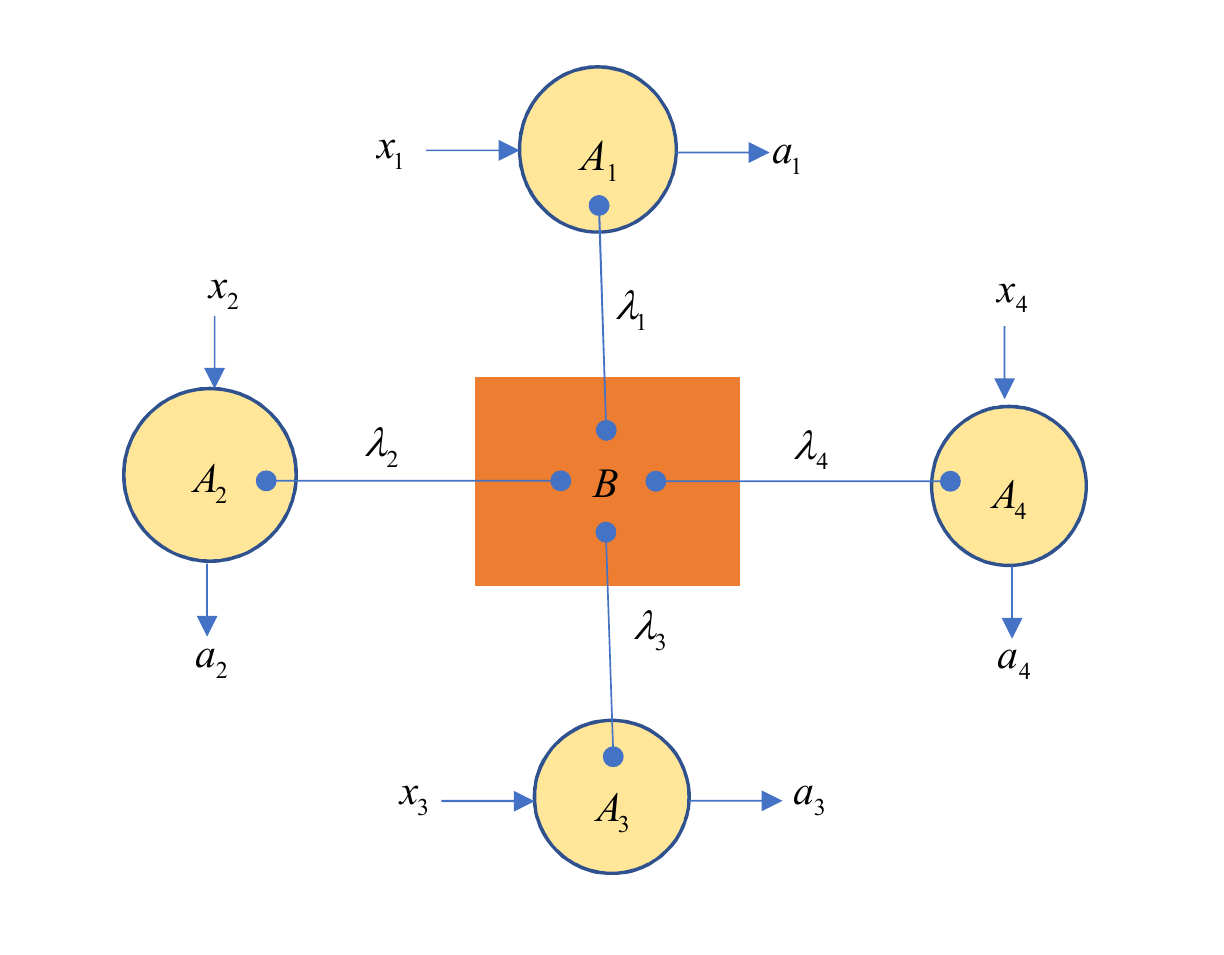}\\
  \caption{Star-network measurement scenario under the $n$-locality assumption where $n=4$.}\label{n-local}
\end{figure}

Consider a star-network measurement scenario Fig. \ref{n-local}, where  Bob and the $i$-th Alice $A_i$ perform simultaneously POVM measurements at the central node $B$ and $i$-th star-node $A_i$, labeled by $y$ and $x_i$, and get their outcomes $b$ and $a_i$, respectively, where $n\ge2$.
The conditional probabilities $P(a_1,\ldots, a_n,b|x_1,\ldots, x_n,y)$ of obtaining results  $a_1,a_2,\ldots,a_n,b$ satisfy
\Bq{CTD}P({\bf{a}}b|{\bf{x}}y)\ge0(\forall {\bf{a}},b,{\bf{x}},y),\ \sum_{{\bf{a}},b}P({\bf{a}}b|{\bf{x}}y)=1(\forall {\bf{x}},y),\Eq
where $n\ge2$, and
$${\bf{a}}=(a_1,\ldots, a_n)\equiv a_1\ldots a_n\in[o_1]\times\ldots\times[o_n],
$$
$$
{\bf{x}}=(x_1,\ldots, x_n)\equiv x_1\ldots x_n\in[m_1]\times\ldots\times[m_n],
$$
and then form an $n+1$-partite CT \cite{BaiLH}:
\Bq{CT1}{\bf{P}}=\Lbrack P(a_1,\ldots, a_n,b|x_1,\ldots, x_n,y)\Rbrack:=\Lbrack P({\bf{a}}b|{\bf{x}}y)\Rbrack\Eq
over
\Bq{Delta}
\Delta_{n+1}=\left(\prod_{i=1}^n[o_i]\right)\times[o_B]\times
\left(\prod_{i=1}^n[m_i]\right)\times[m_B].\Eq

As a generalization of Section 2, this section is devoted to the discussion about $n$-locality of CTs. To do this, we use $\mathcal{CT}(\Delta_{n+1})$ to denote the set of all CTs over $\Delta_{n+1}$ (functions $P$ with condition (\ref{CTD})) and put $N_i=(o_i)^{m_i}(i=1,2,\ldots,n),N_B=(o_B)^{m_B}$. To consider operations of CTs, we use $\mathcal{T}(\Delta_{n+1})$ to denote the set of all tensors (i.e., real functions on $\Delta_{n+1}$) over $\Delta_{n+1}$. For any two elements ${\bf{P}}=\Lbrack P({\bf{a}}b|{\bf{x}}y)\Rbrack$ and ${\bf{Q}}=\Lbrack Q({\bf{a}}b|{\bf{x}}y)\Rbrack$ of  $\mathcal{T}(\Delta_{n+1})$, define
$$s{\bf{P}}+t{\bf{Q}}=\Lbrack sP({\bf{a}}b|{\bf{x}}y)+tQ({\bf{a}}b|{\bf{x}}y)\Rbrack,\ \ \forall s,t\in\R,$$
$$\<{\bf{P}}|{\bf{Q}}\>=\sum_{{\bf{a}},b,{\bf{x}},y} P({\bf{a}}b|{\bf{x}}y)Q({\bf{a}}b|{\bf{x}}y).$$
Then $\mathcal{T}(\Delta_{n+1})$ becomes a real Hilbert space. The elements of $\mathcal{T}(\Delta_{n+1})$ were called correlation-type tensors in \cite{BaiLH}. Clearly, the norm-topology of $\mathcal{T}(\Delta_{n+1})$ is just the pointwise convergent topology on $\Delta_{n+1}$. Thus,
 $\mathcal{CT}(\Delta_{n+1})$ forms a compact convex subset of $\mathcal{T}(\Delta_{n+1})$.

 According to \cite{[21],[23],Tavakoli}, we fix the concept of $n$-locality of a CT over $\Delta_{n+1}$ as follows.

{\bf Definition 4.1.} An $n+1$-partite CT   ${\bf{P}}=\Lbrack P({\bf{a}}b|{\bf{x}}y)\Rbrack$  over $\Delta_{n+1}$ is said to $n$-local if it  admits a continuous $n$-LHVM (C-$n$LHVM):
\begin{equation}\label{nbi-MCT22}
P({\bf{a}}b|{\bf{x}}y)=\int_{\Lambda}\prod_{i=1}^nq_i(\lm_i)\times
\prod_{i=1}^nP_i(a_i|x_i,\lm_i)\times P_B(b|y,\lm){\rm{d}}\mu(\lm)
\end{equation}
for some product measure space $$(\Lambda,\Omega,\mu
)=\left(\Lambda_1\times\ldots\times\Lambda_n,
\Omega_1\times\ldots\times\Omega_n,\mu_1\times\ldots\times\mu_n\right),$$
where $
{\lm}=(\lm_1\lm_2,\ldots, \lm_n)\equiv \lm_1\lm_2\ldots \lm_n\in\Lambda_1\times\ldots\times\Lambda_n$, and

(a) $q_i(\lm_i)$ and $P_i(a_i|x_i,\lm_i)(x_i\in[m_i],a_i\in[o_i])$ are nonnegative $\Omega_i$-measurable functions on $\Lambda_i$,  and $P_B(b|y,\lm)(y\in[m_B],b\in[o_B])$ are nonnegative $\Omega$-measurable functions on $\Lambda$;

(b) $q_i(\lm_i),P_i(a_i|x_i,\lm_i)$ and $P_B(b|y,\lm)$ are PDs of $\lm_i,a_i$ and $b$, respectively, for all $i\in[n]$, $x_i\in[m_i], y\in[m_B]$ and all $\lm\in\Lambda.$

Moreover, ${\bf{P}}$ is said to non-$n$-local if it is not $n$-local.

{\bf Remark 4.1.} Mathematically, $n$-local CTs ${\bf{P}}$ are just ones whose entries $P({\bf{a}}b|{\bf{x}}y)$ can be factorized as a weighted average of the product of $n$ local conditional distributions $P_i(a_i|x_i,\lm_i)$ with parameters $\lm_i(i=1,2,\ldots,n)$ and a conditional distribution $P_B(b|y,\lm)$ with $n$ parameters $\lm_1,\ldots,\lm_n$. For example, when
${\bf{P}}=\Lbrack P({\bf{a}}b|{\bf{x}}y)\Rbrack$ over $\Delta_{n+1}$ is a product CT, i.e.,
$$P({\bf{a}}b|{\bf{x}}y)=P_1(a_1|x_1)\ldots P_n(a_n|x_n)P_B(b|y),$$ it can be written as the form of (\ref{nbi-MCT22}) for the counting measures $\mu_i$ on $\Lambda_i=\{1\}$ and
$$q_i(\lm_i)=1,P_i(a_i|x_i,\lm_i)=P_i(a_i|x_i)(i=1,2,\ldots,n),P_B(b|y,\lm)=P_B(b|y).$$
Thus, every product CT  over $\Delta_{n+1}$ is $n$-local, but not the  converse.

{\bf Remark 4.2.} If an $n+1$-partite CT   ${\bf{P}}=\Lbrack P({\bf{a}}b|{\bf{x}}y)\Rbrack$ over $\Delta_{n+1}$ admits a discrete $n$-LHVM (D-$n$LHVM):
\begin{equation}\label{nDddbi-T222}
P({\bf{a}}b|{\bf{x}}y)=
\sum_{\lm_1=1}^{d_1}\cdots\sum_{\lm_n=1}^{d_n}
\prod_{i=1}^nq_i(\lm_i)\times\prod_{i=1}^nP_i(a_i|x_i,\lm_i)\times P_B(b|y,\lm)
\end{equation}
for all ${\bf{x}},{\bf{a}},y,b,$ where
$$\{q_i(\lm_i)\}_{\lm_i=1}^{n_i},\{P_i(a_i|x_i,\lm_i)\}_{a_i=1}^{o_i}{\rm{\ and\ }} \{P_B(b|y,\lm)\}_{b=1}^{o_B}$$
are PDs, then Eq. (\ref{nbi-MCT22}) holds for the counting measures $\mu_i$ on $\Lambda_i=[d_i](i=1,2,\ldots,n)$. This shows that if ${\bf{P}}$ has a D-$n$LHVM (\ref{nDddbi-T222}), then it has a C-$n$LHVM (\ref{nbi-MCT22}). Indeed, the converse is also valid, see the following theorem.

{\bf Remark 4.3.} Let ${\bf{P}}=\Lbrack P({\bf{a}}b|{\bf{x}}y)\Rbrack$ be  $n$-local. Then it has a C-$n$LHVM (\ref{nbi-MCT22}) and is then  nonsignaling\cite{BaiLH} with the marginal distribution on system $A_iA_jB(i<j)$:
\begin{eqnarray*}
P_{A_iA_jB}(a_ia_jb|x_ix_jy)&=&\iint_{\Lambda_i\times\Lambda_j}
q_i(\lm_i)q_j(\lm_j)
P_i(a_i|x_i,\lm_i)\\
&&\times P_j(a_j|x_j,\lm_j)P_B(b|y,\lm_i\lm_j){\rm{d}}\mu_i(\lm_i){\rm{d}}\mu_j(\lm_j).
\end{eqnarray*}
Thus, ${\bf{P}}_{A_iA_jB}=\Lbrack P_{A_iA_jB}(a_ia_jb|x_ix_jy)\Rbrack$ is a bilocal tripartite CT in the sense of (\ref{C-bi-CT22}) for all $1\le i<j\le n$. Hence, if
there exists  an index $(i_0,j_0)$ with $1\le i_0<j_0\le n$ such that the tripartite CT ${\bf{P}}_{A_{i_0}A_{j_0}B}$ is not bilocal according to definition (\ref{C-bi-CT22}), then  ${\bf{P}}$ must be non-$n$-local. Furthermore, by tracing out part $B$, we get the marginal distribution of ${\bf{P}}$ on system $A_1\cdots A_n$:
$$P_{A_1\cdots A_n}({\bf{a}}|{\bf{x}})=P_{A_1}(a_1|x_1)\cdots P_{A_n}(a_n|x_n), \ \forall{\bf{a}},{\bf{x}},$$
that is, ${\bf{P}}_{A_1\cdots A_n}={\bf{P}}_{A_1}\OO\cdots\OO{\bf{P}}_{A_n},$ a complete product $n$-partite CT. Moreover, when trace out any Alice, say $A_1$, we get

Next, we use $\{J^{(i)}_{k_i}:k_i\in[N_i]\}$ to denote the set of all mappings from $[m_i]$ into $[o_i]$ for all $i=1,2,\ldots,n$ and use $\{K_j:j\in[N_B]\}$ to denote the set of all mappings from $[m_B]$ into $[o_B]$.

{\bf Theorem 4.1.} {\it For a CT ${\bf{P}}=\Lbrack P({\bf{a}}b|{\bf{x}}y)\Rbrack$  over $\Delta_{n+1}$, the following  statements $(i)$-$(v)$ are equivalent.

(i) ${\bf{P}}$ is $n$-local.

(ii) ${\bf{P}}$ has the form:
\begin{equation}\label{nddbi-T222}
P({\bf{a}}b|{\bf{x}}y)=
\sum_{k_1=1}^{N_1}\cdots\sum_{k_n=1}^{N_n}\sum_{j=1}^{N_B}
\pi({k_1,\ldots, k_n,j}) \times\prod_{i=1}^n
\delta_{a_{i},J^{(i)}_{k_i}(x_i)} \times\delta_{b,K_j(y)}
\end{equation}
for all ${\bf{x}},{\bf{a}},y,b,$ where
\Bq{n-PD}
\pi({k_1,\ldots, k_n,j})=\int_{\Lambda}\prod_{i=1}^nq_i(\lm_i)\times
\prod_{i=1}^n\alpha_{k_i}(\lm_i)\times\beta_j(\lm)
{\rm{d}}\mu(\lm),
\Eq
$q_i(\lm_i)$  is PD of $\lm_i$,  $\alpha_{k_i}(\lm_i)$ and  $\beta_j(\lm)$ are PDs of $k_i$ and $j$, respectively, and are measurable w.r.t. $\lm_i$ and $\lm$, respectively.

(iii) ${\bf{P}}$ admits a D-$n$LHVM:
\begin{equation}\label{nddbi-D2222}
P({\bf{a}}b|{\bf{x}}y)=
\sum_{k_1=1}^{N_1}\cdots\sum_{k_n=1}^{N_n}
\prod_{i=1}^n\pi_i(k_i)\times\prod_{i=1}^n
\delta_{a_{i},J^{(i)}_{k_i}(x_i)} \times P_B(b|y,k_1,\ldots,k_n)
\end{equation}
where $\pi_i(k_i)$ and $P_B(b|y,k_1,\ldots,k_n)$ are PDs of $k_i$ and $b$, respectively.

(iv)  ${\bf{P}}$ is ``separable quantum", i.e., it can be generated by a family of local POVMs
$$M^{xy}=\{M_{a_1|x_1}\OO \ldots\OO M_{a_n|x_n}\OO N_{b|y}\}_{(a_1,\ldots,a_n,b)\in[o_1]\times\ldots\times[o_n]\times[o_B]}$$ on a  Hilbert space $\left(\OO_{i=1}^n\H_{A_i}\right)\otimes\H_{B}$ ($\H_B=\OO_{i=1}^n\H_{B_i}$)  together with $n$ separable states  $\rho_{A_iB_i}$ of  systems $\H_{A_i}\otimes\H_{B_i}$, in such a way that
\begin{equation}\label{nQ-bilocal}
P({\bf{a}}b|{\bf{x}}y)=\tr[((\OO_{i=1}^nM_{a_i|x_i})\OO N_{b|y})\rho_{n}]
\end{equation}
for all ${\bf{x}},{\bf{a}},y,b$, where
\Bq{state}
\rho_n=\mathcal{T}(\rho_{A_1B_1}\OO\ldots\OO\rho_{A_nB_n})\mathcal{T}^\dag,
\Eq
and $\mathcal{T}$ denotes   the canonical unitary isomorphism from  $\OO_{i=1}^n(\H_{A_i}\otimes\H_{B_i})$ onto $\left(\OO_{i=1}^n\H_{A_i}\right)\otimes\left(\OO_{i=1}^n\H_{B_i}\right)$, referring to Fig. \ref{starQN} for the case where $n=4$.

(v) {\bf{P}} admits a D-$n$LHVM (\ref{nDddbi-T222}).
}

{\bf Proof.} $(i)\Rightarrow(ii):$ Let $(i)$ be valid. Then the conditions $(a)$ and $(b)$ in Definition 4.1 imply that
$$M_i(\lm_i)=[P_A(a_i|x_i,\lm_i)]_{x_i,a_i}(i=1,2,\ldots,n), M(\lm)=[P_B(b|y,\lm)]_{y,b}$$
are measurable RS function matrices on $\Lambda_i$ and $\Lambda$, respectively. It follows from Lemma 2.1 that they have the following decompositions:
\begin{eqnarray}
M_i(\lm_i)=\sum_{k_i=1}^{N_{i}}\alpha_{k_i}(\lm_i)[\delta_{a_i,J^{(i)}_{k_i}(x_i)}],\ M(\lm)=\sum_{j=1}^{N_B}\beta_j(\lm)[\delta_{b,K_j(y)}],\label{M}
\end{eqnarray}
where $\alpha_{k_i}(\lm_i)$ and $\beta_j(\lm)$ are PDs of $k_i$ and $j$, and are measurable w.r.t. $\lm_i$ and $\lm$, respectively,
$\{R^{(i)}_{k_i}\}_{k_i=1}^{N_{i}}$ and $\{Q_j\}_{j=1}^{N_B}$ denote the sets of all $\{0,1\}$-row-stochastic matrices of $m_i\times o_i$ and $m_B\times o_B$, respectively.
Thus,
$$
P_i(a_i|x_i,\lm_i)=\sum_{k_i=1}^{N_{i}}\alpha_{k_i}(\lm_i)\delta_{a_i,J^{(i)}_{k_i}(x_i)}, P_B(b|y,\lm)=\sum_{j=1}^{N_B}\beta_j(\lm)\delta_{b,K_j(y)},
$$
and   Eqs. (\ref{nddbi-T222}) and (\ref{n-PD}) are then obtained from the C-LHVM (\ref{nbi-MCT22}).

$(ii)\Rightarrow(iii):$
Let statement $(ii)$ be valid. For all $(i,k_i)\in[n]\times[N_i]$, put
$$
\pi_i(k_i)=\int_{\Lambda_i}q_i(\lm_i)\alpha_{k_i}(\lm_i){\rm{d}}\mu_i(\lm_i),
f(b,y,\lm)=\sum_{j=1}^{N_B}\delta_{b,K_j(y)}\beta_j(\lm),$$
$$P_B(b|y,k_1,\ldots,k_n)=\frac{1}{\pi_1({k_1})\ldots\pi_n(k_n)}
\int_{\Lambda}\prod_{i=1}^nq_i(\lm_i)\times\prod_{i=1}^n\alpha_{k_i}(\lm_i)
\times f(b,y,\lm){\rm{d}}\mu(\lm)$$
for all $y\in[m_B],b\in[o_B]$ if $\pi_1({k_1})\ldots\pi_n(k_n)>0$; otherwise, define
$$P_B(b|y,k_1,\ldots,k_n)=\frac{1}{o_B},\ \ \forall y\in[m_B], \forall b\in[o_B].$$
Then $\{\pi_i(k_i)\}_{k_i\in[N_A]}$ and  $\{P_B(b|y,k_1,\ldots,k_n)\}_{b\in[o_B]}$ are PDs, and
$$\int_{\Lambda}\prod_{i=1}^nq_i(\lm_i)\times\prod_{i=1}^n
\alpha_{k_i}(\lm_i)\times f(b,y,\lm){\rm{d}}\mu(\lm)=\prod_{i=1}^n\pi_i(k_i)\times P_B(b|y,k_1,\ldots,k_n)$$
for all $k_i\in[N_i],y\in[m_B],b\in[o_B]$. Then Eq. (\ref{nddbi-T222}) leads to Eq. (\ref{nddbi-D2222}) and so $(iii)$ is valid.

$(iii)\Rightarrow(iv):$ Let $(iii)$ be valid. By putting
$$\H_{A_i}=\H_{B_i}=\C^{N_i},\H_B=\H_{B_1}\OO\H_{B_2}\OO\ldots\OO\H_{B_n},$$
taking orthonormal bases $\{|e^{(i)}_{k_i}\>\}_{k_i=1}^{N_i}$  for $\H_{A_i}=\H_{B_i}$,   defining POVMs:
$$M^{xy}=\{M_{a_1|x_1}\OO \ldots\OO M_{a_n|x_n}\OO N_{b|y}\}_{(a_1,\ldots,a_n,b)\in[o_1]\times\ldots\times[o_n]\times[o_B]}$$ on a  Hilbert space $\OO_{i=1}^n\H_{A_i}\otimes\H_{B}$ with
$$M_{a_i|x_i}=\sum_{k_i=1}^{N_i}\delta_{a_i,J^{(i)}_{k_i}(x_i)}|e^{(i)}_{k_i}\>\<e^{(i)}_{k_i}|,$$
$$N_{b|y}=\sum_{k_1,\ldots,k_n}P_B(b|y,k_1,\ldots,k_n)
\OO_{i=1}^n|e^{(i)}_{k_i}\>\<e^{(i)}_{k_i}|,$$
and constructing separable states:
\Bq{nRAB}\rho_{A_iB_i}=\sum_{k_i=1}^{N_i}\pi_i(k_i)
|e^{(i)}_{k_i}\>_{A_i}\<e^{(i)}_{k_i}|\OO|e^{(i)}_{k_i}\>_{B_i}\<e^{(i)}_{k_i}|,\Eq
we  obtain Eq. (\ref{nQ-bilocal}) from Eq. (\ref{nddbi-D2222}).

$(iv)\Rightarrow(v):$ Let $(iv)$ be valid. Since  $\rho_{A_iB_i}$ is a  separable state of system  $A_iB_i$, it can be written as
$$\rho_{A_iB_i}=\sum_{\lm_i=1}^{d_i}q_i(\lm_i)
|e^{(i)}_{\lm_i}\>\<e^{(i)}_{\lm_i}|\OO|f^{(i)}_{\lm_i}\>\<f^{(i)}_{\lm_i}|,$$
where $\{q_i(\lm_i)\}_{\lm_i\in[d_i]}(i\in[n])$  are PDs,  $\{|e^{(i)}_{\lm_i}\>\}_{\lm_i=1}^{d_i}$ and $\{|f^{(i)}_{\lm_i}\>\}_{\lm_i=1}^{d_i}$ are pure states of $\H_{A_i}$ and $\H_{B_i}$, respectively. Thus,
\begin{eqnarray*}
\rho_n&:=&\mathcal{T}(\rho_{A_1B_1}\OO\ldots\OO\rho_{A_nB_n})\mathcal{T}^\dag\\
&=&\sum_{\lm_1,\ldots,\lm_n}\prod_{i=1}^nq_i(\lm_i)
\left(\OO_{i=1}^n|e^{(i)}_{\lm_i}\>\<e^{(i)}_{\lm_i}|\right)\OO \left(\OO_{i=1}^n|f^{(i)}_{\lm_i}\>\<f^{(i)}_{\lm_i}|\right),
\end{eqnarray*}
and so
\begin{eqnarray*}
P({\bf{a}}b|{\bf{x}}y)&=&\tr[((\OO_{i=1}^nM_{a_i|x_i})\OO N_{b|y})\rho_{n}]\\
&=&\sum_{\lm_1,\ldots,\lm_n}\prod_{i=1}^nq_i(\lm_i)\times\prod_{i=1}^nP_i(a_i|x_i,\lm_i)\times P_B(b|y,\lm_1,\ldots,\lm_n)
\end{eqnarray*}
for all ${\bf{x}},{\bf{a}},y,b$, where
$$P_i(a_i|x_i,\lm_i)=\left\<e^{(i)}_{\lm_i}|M_{a_i|x_i}|e^{(i)}_{\lm_i}\right\>,
$$$$P_B(b|y,\lm_1,\ldots,\lm_n)=\left\<f^{(1)}_{\lm_1}\ldots f^{(n)}_{\lm_n}|N_{b|y}|
f^{(1)}_{\lm_1}\ldots f^{(n)}_{\lm_n}\right\>.$$
This shows that ${\bf{P}}$ has a D-$n$LVHM.

$(v)\Rightarrow(i):$ Use Remark 4.2.
The proof is completed.

As an application of Theorem 4.1, we obtain the following corollary.

{\bf{Corollary 4.1.}} {\it A CT ${\bf{P}}=\Lbrack P({\bf{a}}b|{\bf{x}}y)\Rbrack$ over $\Delta_{n+1}$ has a  C-$n$LHVM (\ref{nbi-MCT22}) if and only if it has a D-$n$LHVM (\ref{nDddbi-T222}) if and only if it can be written as
\begin{equation}\label{2}
P({\bf{a}}b|{\bf{x}}y)=
\sum_{k_1=1}^{N_1}\cdots\sum_{k_n=1}^{N_n}\sum_{j=1}^{N_B}q(k_1,\ldots,k_n,j)
\prod_{i=1}^n
\delta_{a_{i},J^{(i)}_{k_i}(x_i)} \times\delta_{b,K_j(y)}
\end{equation}
for all ${\bf{x}},{\bf{a}},y,b,$ where
\Bq{q}q(k_1,\ldots,k_n,j)=\pi_{1}(k_1)\ldots\pi_{n}(k_n)p(j|k_1,\ldots,k_n),\Eq
$\pi_{i}(k_i)$ and $p(j|k_1,\ldots,k_n)$ are PDs of $k_i$ and $j$, respectively.}

From this corollary, we see that when a CT ${\bf{P}}=\Lbrack P({\bf{a}}b|{\bf{x}}y)\Rbrack$ over $\Delta_{n+1}$
is $n$-local, it has  a D-$n$LHVM (\ref{nDddbi-T222}). Tracing out any Alice, say $A_1$, yields
\begin{eqnarray}\label{nDddbi-T2223}
P(a_2\ldots a_nb|x_2\ldots x_ny)&=&
\sum_{\lm_2=1}^{d_2}\ldots\sum_{\lm_n=1}^{d_n}
\prod_{i=2}^nq_i(\lm_i)\nonumber\\&&\times \prod_{i=2}^nP_i(a_i|x_i,\lm_i)\times P'_B(b|y,\lm_2\ldots \lm_n)
\end{eqnarray}
for all ${a_2,\ldots,a_n},{x_2,\ldots,x_n},y,b,$ where
$$P'_B(b|y,\lm_2\cdots \lm_n)=\sum_{\lm_1=1}^{d_1}q_i(\lm_i)P_B(b|y,\lm),$$ which is a PD of $b$ for all $y,\lm_2\cdots \lm_n$. This means that the marginal distribution ${\bf{P}}_{A_2\cdots A_nB}$ of an $n$-local CT ${\bf{P}}$ is an $n-1$-local CT.

It was proved in \cite[Theorem 5.1]{BaiLH} that an $n+1$-partite  CT ${\bf{P}}=\Lbrack P({\bf{a}}b|{\bf{x}}y)\Rbrack$  over $\Delta_{n+1}$ is Bell local if and only if Eq. (\ref{2}) holds for some PD $q(k_1,\ldots,k_n,j)$ of $k_1,\ldots,k_n,j$, which is not necessarily of the form (\ref{q}). This characterization implies that
the set $\mathcal{CT}^{\textrm{Bell-local}}(\Delta_{n+1})$  of all $n+1$-partite Bell local CTs  over $\Delta_{n+1}$ forms a convex compact set.
Thus, Corollary 2.1 implies that every $n$-local CT ${\bf{P}}=\Lbrack P({\bf{a}}b|{\bf{x}}y)\Rbrack$  over $\Delta_{n+1}$  must be $n+1$-partite Bell local CT. It follows from Remark 4.1 that the set $\mathcal{CT}^{\textrm{Bell-local}}(\Delta_{n+1})$  is just the convex hull of the set
 $\mathcal{CT}^{n\textrm{-local}}(\Delta_{n+1})$ of all $n$-local CTs over $\Delta_{n+1}$. That is,
 \Bq{lnL}\mathcal{CT}^{\textrm{Bell-local}}(\Delta_{n+1})={\rm{conv}}(\mathcal{CT}^{n\textrm{-local}}(\Delta_{n+1})).\Eq

 The  characterization (\ref{nddbi-D2222}) of an $n$-local CT shows that all $n$-local CTs can be represented by D-$n$LHVMs in which local hidden variables $k_1,k_2,\ldots,k_n$ have the same dimensions $N_1,N_2,\ldots,N_n$. This advantage will serve to the proof of the closedness of $\mathcal{CT}^{n\textrm{-local}}(\Delta_{n+1})$.

{\bf{Corollary 4.2.}} {\it $\mathcal{CT}^{n\textrm{-local}}(\Delta_{n+1})$ is a compact set in the Hilbert space $\mathcal{T}(\Delta_{n+1})$.}

{\bf{Proof.}} Let ${\bf{P}}^{(m)}=\Lbrack P^{(m)}({\bf{a}}b|{\bf{x}}y)\Rbrack \in\mathcal{CT}^{n\textrm{-local}}(\Delta_{n+1})$ for all $m=1,2,\ldots,$ with $\textbf{P}^{(m)}\rightarrow \textbf{P}\in\mathcal{T}(\Delta_{n+1})$ as $m\rightarrow\infty$, i.e., $P^{(m)}({\bf{a}}b|{\bf{x}}y)\rightarrow P({\bf{a}}b|{\bf{x}}y)$  as $m\rightarrow\infty$ for all possible variables $({\bf{a}},b,{\bf{x}},y)$. Clearly, $\textbf{P}\in\mathcal{CT}(\Delta_{n+1})$. According to Theorem 4.1, each $\textbf{P}^{(m)}$ can be written as
\begin{equation}\label{mnddbi-D222}
P^{(m)}({\bf{a}}b|{\bf{x}}y)=
\sum_{k_1=1}^{N_1}\cdots\sum_{k_n=1}^{N_n}
\prod_{i=1}^n\pi^{(m)}_i(k_i)\times\prod_{i=1}^n
\delta_{a_{i},J^{(i)}_{k_i}(x_i)} \times P^{(m)}_B(b|y,k_1\ldots k_n)
\end{equation}
for all $m=1,2,\ldots$ and   possible variables $({\bf{a}},b,{\bf{x}},y)$, where $\{\pi^{(m)}_i(k_i)\}_{k_i\in[N_i]}(\forall i\in[n])$ and $\{P^{(m)}_B(b|y,k_1\ldots k_n)\}_{b\in[o_B]}(\forall y\in[m_B],k_i\in[N_i])$ are PDs.
By taking subsequences if necessary, we may assume that
$$\lim_{m\rightarrow\infty}\pi^{(m)}_i(k_i)=\pi_i(k_i)(\forall i\in[n]),$$
$$\lim_{m\rightarrow\infty}P^{(m)}_B(b|y,k_1\ldots k_n)=P_B(b|y, k_1\ldots k_n)(\forall y,b,k_1,\ldots, k_n).$$
Obviously, $\{\pi_i(k_i)\}_{i\in[N_i]}$ and   $\{P_B(b|y,k_1\ldots k_n)\}_{b\in[o_B]}(\forall y,k_1,\ldots,k_n)$ are PDs. Letting $m\rightarrow\infty$ in Eq. (\ref{mnddbi-D222})  yields that
$$
P({\bf{a}}b|{\bf{x}}y)=
\sum_{k_1=1}^{N_1}\cdots\sum_{k_n=1}^{N_n}
\prod_{i=1}^n\pi_i(k_i)\times\prod_{i=1}^n
\delta_{a_{i},J^{(i)}_{k_i}(x_i)} \times P_B(b|y,k_1\ldots k_n)
$$
for all possible ${\bf{x}},{\bf{a}},y,b$. Using Theorem 4.1 again implies that
${\bf{P}}\in\mathcal{CT}^{n\textrm{-local}}(\Delta_{n+1})$. This shows that $\mathcal{CT}^{n\textrm{-local}}(\Delta_{n+1})$ is closed and so compact since it is also a bounded set of the finite dimensional Hilbert space $\mathcal{T}(\Delta_{n+1})$.   The proof is completed.

From Definition 4.1, we observe  that when
${\bf{P}}\in\mathcal{CT}^{n\textrm{-local}}(\Delta_{n+1})$, the marginal distributions satisfy $
{\bf{P}}_{A_1\ldots A_n}={\bf{P}}_{A_1}\OO\ldots\OO{\bf{P}}_{A_n},$
i.e.,
$${{P}}_{A_1\ldots A_n}(a_1\ldots a_n|x_1\ldots x_n)={{P}}_{A_1}(a_1|x_1)\ldots{{P}}_{A_n}(a_n|x_n),\ \forall a_i,x_i;$$
particularly,
\Bq{P-product2}
{\bf{P}}_{A_1A_2}={\bf{P}}_{A_1}\OO{\bf{P}}_{A_2}.
\Eq
It is easy to find  a Bell local ${\bf{P}}$ such that the property  (\ref{P-product2}) is not satisfied and then is not $n$-local. This shows that not all Bell local CTs are $n$-local. By noticing that a convex combination of two product CTs is not necessarily a product CT, we conclude from the property  (\ref{P-product2}) that  the set $\mathcal{CT}^{n\textrm{-local}}(\Delta_{n+1})$ is not convex. However, it has many subsets that are star-convex.

Next, let us discuss the weak star-convexity of the set $\mathcal{CT}^{n\textrm{-local}}(\Delta_{n+1})$ by finding star-convex subsets of it. To do so, for a fixed $1\le k\le n$, put
$${\widehat{A_k}}=A_1\ldots A_{k-1}A_{k+1}\ldots A_{n},$$
$${\bf{a}}_k=a_1\ldots a_{k-1}a_{k+1}\ldots a_n,
{\bf{x}}_k=x_1\ldots x_{k-1}x_{k+1}\ldots x_n,$$
and take an $n-1$-partite product CT ${\bf{E}}_k=\Lbrack E_k({\bf{a}}_k|{\bf{x}}_k)\Rbrack$ over $\prod_{i\ne k}[o_i]\times
\prod_{i\ne k}[m_i]$  with
$$E_k({\bf{a}}_k|{\bf{x}}_k)=\prod_{i\ne k}Q_i(a_i|x_i),$$ and define a set
\Bq{CT-Ek}
\mathcal{CT}_{{\bf{E}}_k}^{n\textrm{-local}}(\Delta_{n+1})=
\left\{{\bf{P}}\in \mathcal{CT}^{n\textrm{-local}}(\Delta_{n+1}): {\bf{P}}_{\widehat{A_k}}={\bf{E}}_k\right\},
\Eq
where ${\bf{P}}_{\widehat{A_k}}$ denotes the marginal distribution of ${\bf{P}}$ on the subsystem $\widehat{A_k}$, i.e.,
$${{P}}_{\widehat{A_k}}({\bf{a}}_k|{\bf{x}}_k)=\sum_{a_k,b}P({\bf{a}}b|{\bf{x}}b),$$
which is independent of the choices of $x_k$ and $y$ whenever ${\bf{P}}$ is $n$-local and then nonsignaling. Define ${\bf{S}}_k=\Lbrack S_k({\bf{a}}b|{\bf{x}}y)\Rbrack$ by
$S_k({\bf{a}}b|{\bf{x}}y)=\prod_{i=1}^nQ_i(a_i|x_i)
\times\frac{1}{o_B},$
where $Q_k(a_k|x_k)\equiv \frac{1}{o_k},$ then ${\bf{S}}_k\in \mathcal{CT}_{{\bf{E}}_k}^{n\textrm{-local}}(\Delta_{n+1})$ (Remark 4.1).

\begin{figure}[h]
  \centering
  \includegraphics[width=10cm]{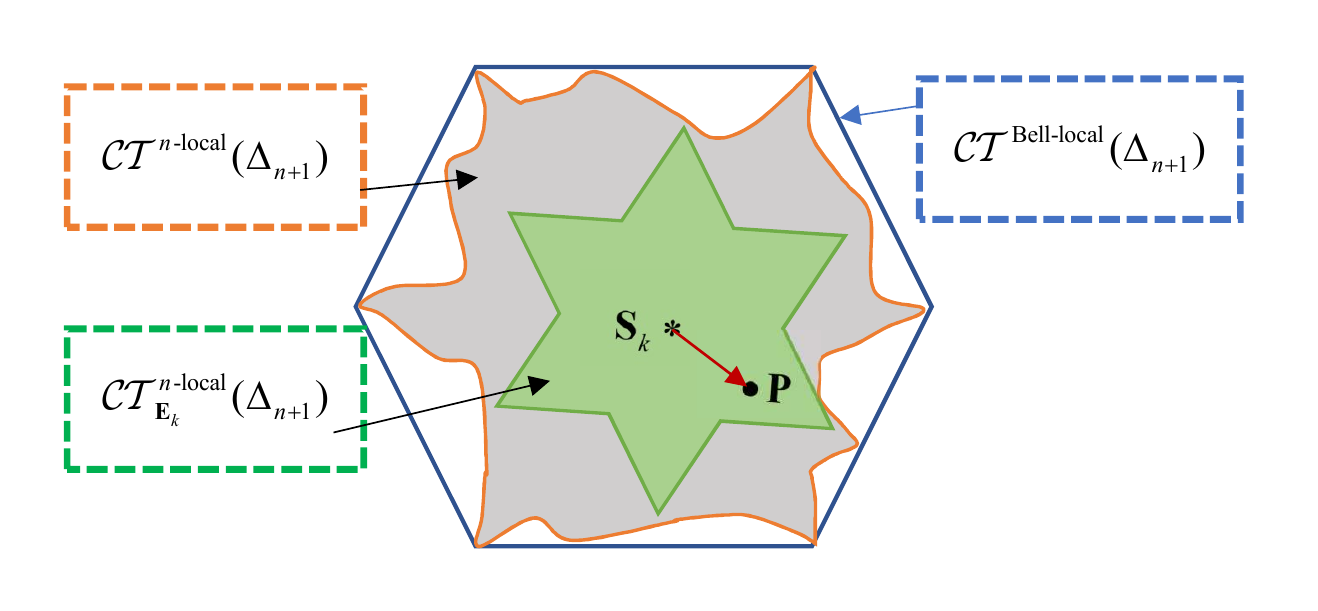}\\
  \caption{Star-convexity of the set $ \mathcal{CT}_{{\bf{E}}_k}^{n\textrm{-local}}(\Delta_{n+1})$.}\label{n-star-convex}
\end{figure}

{\bf{Corollary  3.3.}} {\it For any $k=1,2,\ldots,n,$ the set} $\mathcal{CT}_{{\bf{E}}_k}^{n\textrm{-local}}(\Delta_{n+1})$  {\it is star-convex with a sun ${\bf{S}}_k$, i.e.,}
$$(1-t){\bf{S}}_k+t \mathcal{CT}_{{\bf{E}}_k}^{n\textrm{-local}}(\Delta_{n+1})
\subset\mathcal{CT}_{{\bf{E}}_k}^{n\textrm{-local}}(\Delta_{n+1}),\ \forall t\in[0,1].$$ {\it See Figure \ref{n-star-convex}.}

\quad{\bf{Proof.}} Let ${\bf{P}}=\Lbrack P({\bf{a}}b|{\bf{x}}y)\Rbrack$ be any element of $\mathcal{CT}_{{\bf{E}}_k}^{n\textrm{-local}}(\Delta_{n+1})$. Then Theorem 4.1 implies that ${\bf{P}}$ has a D-$n$LHVM:
\begin{equation}\label{nD-T}
P({\bf{a}}b|{\bf{x}}y)=
\sum_{\lm_1=1}^{d_1}\cdots\sum_{\lm_n=1}^{d_n}
\prod_{i=1}^nq_i(\lm_i)\times\prod_{i=1}^nP_i(a_i|x_i,\lm_i)\times P_B(b|y,\lm)
\end{equation}
for all ${\bf{x}},{\bf{a}},y,b,$ where
$$\{q_i(\lm_i)\}_{\lm_i=1}^{n_i},\{P_i(a_i|x_i,\lm_i)\}_{a_i=1}^{o_i}{\rm{\ and\ }} \{P_B(b|y,\lm)\}_{b=1}^{o_B}$$
are PDs. For every $t\in[0,1]$, put
\Bq{nfs}
f_k(\lm_k,s)=\left\{\begin{array}{cc}
q_k(\lm_k)(1-t), & s=0; \\
                    q_k(\lm_k)t, & s=1,
                  \end{array}\right.
\Eq
\Bq{npbs}
P_B(b|y,\lm_1\ldots\lm_{k-1}(\lm_k,s)\lm_{k+1}\ldots\lm_n)
=\left\{\begin{array}{cc}
                    \frac{1}{o_B}, & s=0; \\
                    P_B(b|y,\lm_1\ldots\lm_{n}), & s=1,
                  \end{array}\right.
\Eq
\Bq{pcs}
P_k(a_k|x_k,(\lm_k,s))=\left\{\begin{array}{cc}
                    \frac{1}{o_k}, & s=0; \\
                    P_k(a_k|x_k,\lm_k), & s=1,
                  \end{array}\right.
\Eq
which are PDs of $(\lm_k,s),b$ and $a_k$, respectively. Put
\begin{eqnarray*}
Q^t_k({\bf{a}}b|{\bf{x}}y)
&=&\sum_{\lm_i(i\ne k)}\prod_{i\ne k}q_i(\lm_i)\times  \sum_{\lm_k,s}f_k(\lm_k,s)\times \prod_{i\ne k}P_i(a_i|x_i,\lm_i)\\ &&\times P_k(a_k|x_k,(\lm_k,s))P_B(b|y,\lm_1\ldots\lm_{k-1}(\lm_k,s)\lm_{k+1}\ldots\lm_n),
\end{eqnarray*}
then Theorem 4.1 implies that ${\bf{Q}}^t_k=\Lbrack Q^t_k({\bf{a}}b|{\bf{x}}y)\Rbrack\in \mathcal{CT}^{n\textrm{-local}}(\Delta_{n+1})$. On the other hand, for   all ${\bf{x}},{\bf{a}},y,b,$  we compute that
\begin{eqnarray*}
&&Q^t_k({\bf{a}}b|{\bf{x}}y)\\
&=&\sum_{\lm_i(i\ne k)}\prod_{i\ne k}q_i(\lm_i)\times \sum_{\lm_k}
f_k(\lm_k,0)\times\prod_{i\ne k}P_i(a_i|x_i,\lm_i)\\ && \times P_k(a_k|x_k,(\lm_k,0))P_B(b|,y\lm_1\ldots\lm_{k-1}(\lm_k,0)\lm_{k+1}\ldots\lm_n)\\
&&+\sum_{\lm_i(i\ne k)}\prod_{i\ne k}q_i(\lm_i)\sum_{\lm_k}
f_k(\lm_k,1)\times\prod_{i\ne k}P_i(a_i|x_i,\lm_i)\\ && \times P_k(a_k|x_k,(\lm_k,1))P_B(b|,y\lm_1\ldots\lm_{k-1}(\lm_k,1)\lm_{k+1}\ldots\lm_n)\\
&=&(1-t)P_{\widehat{A_k}}({\bf{a}}_k|{\bf{x}}_k)\times
\frac{1}{o_k}\times\frac{1}{o_B}+tP({\bf{a}}b|{\bf{x}}y)\\
&=&(1-t)S_k({\bf{a}}b|{\bf{x}}y)+tP({\bf{a}}b|{\bf{x}}y).
\end{eqnarray*}
This shows that $(1-t){\bf{S}}_k+t{\bf{P}}={\bf{Q}}^t_k$, which is an $n$-local CT  over $\Delta_{n+1}$. Clearly,
$$({\bf{Q}}^t_k)_{\widehat{A_k}}=(1-t)({\bf{S}}_k)_{\widehat{A_k}}+t{\bf{P}}_{\widehat{A_k}}
={\bf{E}}_k$$ and so
${\bf{Q}}^t_k$, i.e., $(1-t){\bf{S}}_k+t{\bf{P}}$ is an element of $\mathcal{CT}_{{\bf{E}}_k}^{n\textrm{-local}}(\Delta_{n+1}).$
This shows that
$$(1-t){\bf{S}}_k+t\mathcal{CT}_{{\bf{E}}_k}^{n\textrm{-local}}(\Delta_{n+1})\subset \mathcal{CT}_{{\bf{E}}_k}^{n\textrm{-local}}(\Delta_{n+1}),\ \ \forall t\in[0,1],$$
and so $\mathcal{CT}_{{\bf{E}}_k}^{n\textrm{-local}}(\Delta_{n+1})$ is star-convex with a sun ${\bf{S}}_k$. The proof is completed.

\begin{figure}[h]
  \centering
  \includegraphics[width=10cm]{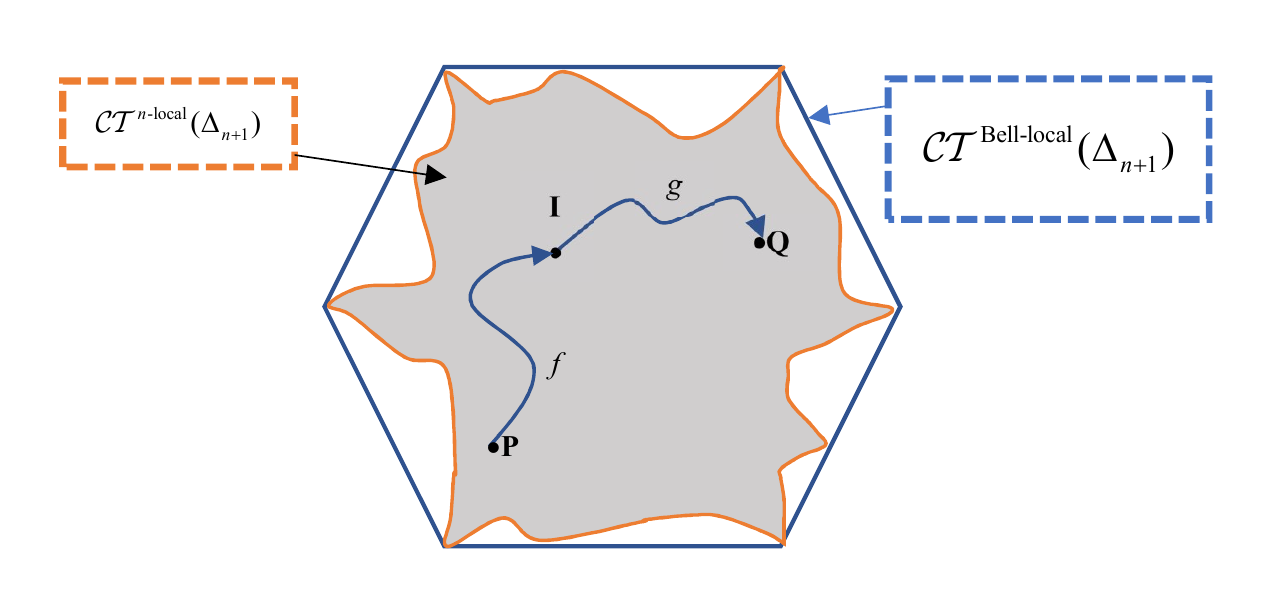}\\
  \caption{Path-connectedness of the set $\mathcal{CT}^{n\textrm{-local}}(\Delta_{n+1})$.}\label{path-connectness}
\end{figure}

{\bf{Corollary 4.4.}} {\it The set} $\mathcal{CT}^{n\textrm{-local}}(\Delta_{n+1})$ {\it is path-connected. See Figure \ref{path-connectness}.}

{\bf Proof.} Put $I({\bf{a}}b|{\bf{x}}y)\equiv\frac{1}{o_1o_2\ldots o_no_B}$, then
${\bf{I}}:=\Lbrack I({\bf{a}}b|{\bf{x}}y)\Rbrack$ is an element of $\mathcal{CT}^{n\textrm{-local}}(\Delta_{n+1})$.
Let ${\bf{P}}=\Lbrack P({\bf{a}}b|{\bf{x}}y)\Rbrack$ and
${\bf{Q}}=\Lbrack Q({\bf{a}}b|{\bf{x}}y)\Rbrack$ be any two elements of
$\mathcal{CT}^{n\textrm{-local}}(\Delta_{n+1})$. Theorem 4.1 implies that  ${\bf{P}}$ and ${\bf{Q}}$ admit D-$n$LHVMs:
\begin{equation}\label{np}
P({\bf{a}}b|{\bf{x}}y)=
\sum_{\lm_1=1}^{d_1}\cdots\sum_{\lm_n=1}^{d_n}
\prod_{i=1}^np_i(\lm_i)\times\prod_{i=1}^nP_i(a_i|x_i,\lm_i)\times P_B(b|y,\lm),
\end{equation}
\begin{equation}\label{np}
Q({\bf{a}}b|{\bf{x}}y)=
\sum_{\xi_1=1}^{r_1}\cdots\sum_{\xi_n=1}^{r_n}
\prod_{i=1}^nq_i(\xi_i)\times\prod_{i=1}^nP_i(a_i|x_i,\xi_i)\times P_B(b|y,\xi)
\end{equation}
where $\xi=\xi_1\xi_2\ldots\xi_n$.
For every $t\in[0,1/2]$, set
$$P^t_i(a_i|x_i,\lm_i)=(1-2t)P_i(a_i|x_i,\lm_i)+2t\frac{1}{o_i}(i=1,2,\ldots,n);$$
$$P^t_B(b|y,\lm)=(1-2t)P_B(b|y,\lm)+2t\frac{1}{o_B},$$
which are clearly PDs of $a_i$ and $b$, respectively. Put
\begin{eqnarray*}
P^t({\bf{a}}b|{\bf{x}}y)=\sum_{\lm_1=1}^{d_1}\cdots\sum_{\lm_n=1}^{d_n} \prod_{i=1}^np_i(\lm_i)\times\prod_{i=1}^nP^t_i(a_i|x_i,\lm_i)\times P^t_B(b|y,\lm),
\end{eqnarray*}
then
${f}(t):=\Lbrack P^t({\bf{a}}b|{\bf{x}}y)\Rbrack$ is an $n$-local CT over $\Delta_{n+1}$ for all $t\in[0,1/2]$ with ${f}(0)={\bf{P}}$ and ${f}(1/2)={\bf{I}}$. Obviously, the map $t\mapsto {f}(t)$ from $[0,1/2]$ into $\mathcal{CT}^{n\textrm{-local}}(\Delta_{n+1})$ is continuous.

Similarly, for every $t\in[1/2,1]$, set
$$Q^t_i(a_i|x_i,\xi_i)=(2t-1)Q_i(a_i|x_i,\xi_i)+2(1-t)\frac{1}{o_i}(i=1,2,\ldots,n),$$
$$Q^t_B(b|y,\xi)=(2t-1)Q_B(b|y,\xi)+2(1-t)\frac{1}{o_B},$$
which are clearly PDs of $a_i$ and $b$, respectively. Put
\begin{eqnarray*}
Q^t({\bf{a}}b|{\bf{x}}y)=\sum_{\xi_1=1}^{r_1}\cdots\sum_{\xi_n=1}^{r_n} \prod_{i=1}^nq_i(\xi_i)\times\prod_{i=1}^nQ^t_i(a_i|x_i,\xi_i)\times P^t_B(b|y,\xi),
\end{eqnarray*}
then
${g}(t):=\Lbrack Q^t({\bf{a}}b|{\bf{x}}y)\Rbrack$ is an $n$-local CT over $\Delta_{n+1}$ for all  $t\in[1/2,1]$ with ${g}(1/2)={\bf{I}}$ and ${g}(1)={\bf{Q}}$. Obviously, the map $t\mapsto {g}(t)$ from $[1/2,1]$ into $\mathcal{CT}^{n\textrm{-local}}(\Delta_{n+1})$ is continuous. Thus, the function
$p:[0,1]\rightarrow \mathcal{CT}^{n\textrm{-local}}(\Delta_{n+1})$ defined by
$$p(t)=\left\{\begin{array}{cc}
                {f}(t), & t\in[0,1/2];\\
                {g}(t), & t\in(1/2,1],
              \end{array}\right.$$
is continuous everywhere
 and then induces a path $p$ in $\mathcal{CT}^{n\textrm{-local}}(\Delta_{n+1})$ with $p(0)={\bf{P}}$ and $p(1)={\bf{Q}}$. This shows that $\mathcal{CT}^{n\textrm{-local}}(\Delta_{n+1})$ is path-connected.  The proof is completed.

\section{$n$-Locality of $n+1$-partite PTs}
\setcounter{equation}{0}\setcounter{section}{5}

In this section, we introduce and discuss $n+1$-partite PT ${\bf{P}}=\Lbrack P(a_1,\ldots, a_n,b):=\Lbrack P({\bf{a}},b)\Rbrack\equiv\Lbrack P({\bf{a}}b)\Rbrack$ over
$O_{n+1}=\prod_{i=1}^n[o_i]\times[o_B]$ where
$${\bf{a}}=(a_1,\ldots, a_n)\equiv a_1\ldots a_n\in[o_1]\times\ldots\times[o_n].
$$ It
is defined as  a tensor with index set $O_{n+1}$, equivalently, a function $P:O_{n+1}\rightarrow\R$, satisfying
\Bq{nPTD}P({\bf{a}},b)\ge0(\forall {\bf{a}},b),\ \sum_{{\bf{a}},b}P({\bf{a}},b)=1.\Eq
We use $\mathcal{PT}(O_{n+1})$ to denote the set of all PTs over $O_{n+1}$ (functions $P$ with condition (\ref{nPTD})).

To consider operations of PTs, we use $\mathcal{T}(O_{n+1})$ to denote the set of all tensors (i.e., real functions) over $O_{n+1}$. For any two elements ${\bf{P}}=\Lbrack P({\bf{a}},b)\Rbrack$ and ${\bf{Q}}=\Lbrack Q({\bf{a}},b)\Rbrack$ of  $\mathcal{T}(O_{n+1})$, define
$$s{\bf{P}}+t{\bf{Q}}=\Lbrack sP({\bf{a}},b)+tQ({\bf{a}},b)\Rbrack,\ \<{\bf{P}}|{\bf{Q}}\>=\sum_{{\bf{a}},b} P({\bf{a}},b)Q({\bf{a}},b).$$
Then $\mathcal{T}(O_{n+1})$ becomes a real Hilbert space.  Clearly, the norm-topology of $\mathcal{T}(O_{n+1})$ is just the pointwise convergent topology on $O_{n+1}$. Thus,
 $\mathcal{PT}(O_{n+1})$ forms a compact convex subset of $\mathcal{T}(O_{n+1})$.

{\bf Definition 5.1.} An $n+1$-partite PT   ${\bf{P}}=\Lbrack P({\bf{a}},b)\Rbrack$  over $\Delta_{n+1}$ is said to $n$-bilocal if it  admits a continuous $n$-LHVM (C-$n$LHVM):
\begin{equation}\label{C-n-PLHVM}
P({\bf{a}},b)=\int_{\Lambda}\prod_{i=1}^nq_i(\lm_i)\times
\prod_{i=1}^nP_i(a_i|\lm_i)\times P_B(b|\lm){\rm{d}}\mu(\lm)
\end{equation}
for some product measure space $$(\Lambda,\Omega,\mu
)=\left(\Lambda_1\times\ldots\times\Lambda_n,
\Omega_1\times\ldots\times\Omega_n,\mu_1\times\ldots\times\mu_n
\right),$$
 where
$
{\lm}=(\lm_1\lm_2,\ldots, \lm_n)\equiv \lm_1\lm_2\ldots \lm_n\in\Lambda_1\times\ldots\times\Lambda_n,
$ and

(a) $q_i(\lm_i)$ and $P_i(a_i|\lm_i)(a_i\in[o_i])$ are nonnegative $\Omega_i$-measurable functions on $\Lambda_i$,  and $P_B(b|\lm)(b\in[o_B])$ are nonnegative $\Omega$-measurable functions on $\Lambda$;

(b) $q_i(\lm_i),P_i(a_i|\lm_i)$ and $P_B(b|\lm)$ are PDs of $\lm_i,a_i$ and $b$, respectively, for all $i\in[n]$ and all $\lm\in\Lambda$.

{\bf Remark 5.1.} If an $n+1$-partite PT   ${\bf{P}}=\Lbrack P({\bf{a}},b)\Rbrack$ over $O_{n+1}$ is product, i.e., $P({\bf{a}}b|{\bf{x}}y)$ is a product of $n+1$ CTs $P_1(a_1),\ldots,P_n(a_n)$ and $P_B(b)$, then it can be written as the form of (\ref{nbi-MCT22}) by taking the counting measures $\mu_i$ on $\Lambda_i=\{1\}$ and
$$q_i(\lm_i)=1,P_1(a_i|\lm_i)=P_i(a_i)(i=1,2,\ldots,n),P_B(b|\lm)=P_B(b).$$
Thus, every product PT  over $O_{n+1}$ is $n$-local, but not the  converse.

{\bf Remark 5.2.} If an $n+1$-partite PT   ${\bf{P}}=\Lbrack P({\bf{a}},b)\Rbrack$ over $O_{n+1}$ admits a discrete $n$-LHVM (D-$n$LHVM):
\begin{equation}\label{D-n-PLHVM}
P({\bf{a}},b)=
\sum_{\lm_1=1}^{d_1}\cdots\sum_{\lm_n=1}^{d_n}
\prod_{i=1}^nq_i(\lm_i)\times\prod_{i=1}^nP_i(a_i|\lm_i)\times P_B(b|\lm)
\end{equation}
for all ${\bf{a}},b,$ where $q_i(\lm_i),P_i(a_i|\lm_i)$ and $P_B(b|\lm)$ are PDs of $\lm_i,a_i$ and $b$, respectively, for all $i\in[n]$ and all $\lm\in\Lambda$, then Eq. (\ref{C-n-PLHVM}) holds for the counting measures $\mu_i$ on $\Lambda_i=[d_i](i=1,2,\ldots,n)$. This shows that if ${\bf{P}}$ has a D-$n$LHVM (\ref{D-n-PLHVM}), then it has a C-$n$LHVM (\ref{C-n-PLHVM}). Indeed, the converse is also valid, see the following theorem which can be viewed as a special case of Theorem 4.1 with $m_i=1$ and $N_i=o_i$ for all $i=1,2,\ldots,n$.

{\bf Theorem 5.1.} {\it For a PT ${\bf{P}}=\Lbrack P({\bf{a}},b)\Rbrack$  over $O_{n+1}$, the following  statements $(i)$-$(iv)$ are equivalent.

(i) ${\bf{P}}$ is $n$-local, i.e., it can be written as (\ref{C-n-PLHVM}).

(ii) ${\bf{P}}$ admits a D-$n$LHVM:
\begin{equation}\label{Pnddbi-D2222}
P({\bf{a}},b)=
\sum_{k_1=1}^{N_1}\cdots\sum_{k_n=1}^{N_n}
\prod_{i=1}^n\pi_i(k_i)\times\prod_{i=1}^n
\delta_{a_{i},{k_i}} \times P_B(b|k_1,\ldots,k_n)
\end{equation}
where $\{\pi_i(k_i)\}_{k_i\in[N_i]}(\forall i\in[n])$ and $\{P_B(b|k_1,\ldots,k_n)\}_{b\in[o_B]}(\forall k_i\in[N_i])$ are PDs.

(iii)  ${\bf{P}}$ is ``separable quantum", i.e., it can be generated by a  local POVM
$$M=\{M_{a_1}\OO \ldots\OO M_{a_n}\OO N_{b}\}_{(a_1,\ldots,a_n,b)\in[o_1]\times\ldots\times[o_n]\times[o_B]}$$ on a  Hilbert space $\left(\OO_{i=1}^n\H_{A_i}\right)\otimes\H_{B}$ ($\H_B=\OO_{i=1}^n\H_{B_i}$)  together with $n$ separable states  $\rho_{A_iB_i}$ of  systems $\H_{A_i}\otimes\H_{B_i}$, in such a way that
\begin{equation}\label{PnQ-bilocal}
P({\bf{a}},b)=\tr[((\OO_{i=1}^nM_{a_i})\OO N_{b})\rho_{n}],\ \ \forall {\bf{a}},b,
\end{equation}
 where
\Bq{state}
\rho_n=\mathcal{T}(\rho_{A_1B_1}\OO\ldots\OO\rho_{A_nB_n})\mathcal{T}^\dag,
\Eq
and $\mathcal{T}$ denotes   the canonical unitary isomorphism from  $\OO_{i=1}^n(\H_{A_i}\otimes\H_{B_i})$ onto $\left(\OO_{i=1}^n\H_{A_i}\right)\otimes\left(\OO_{i=1}^n\H_{B_i}\right)$.

(iv) {\bf{P}} admits a D-$n$LHVM:
\begin{equation}\label{ppD-n-PLHVM}
P({\bf{a}},b)=
\sum_{\lm_1=1}^{d_1}\cdots\sum_{\lm_n=1}^{d_n}
\prod_{i=1}^nq_i(\lm_i)\times\prod_{i=1}^nP_i(a_i|\lm_i)\times P_B(b|\lm),\ \ \forall {\bf{a}},b,
\end{equation}
where $\{q_i(\lm_i)\}_{\lm_i=1}^{n_i},\{P_i(a_i|\lm_i)\}_{a_i=1}^{o_i}{\rm{\ and\ }} \{P_B(b|\lm)\}_{b=1}^{o_B}$
are PDs.}

As special cases of corresponding conclusions in Section 4, we obtain the following corollaries.

{\bf{Corollary 5.1.}} {\it A PT  over $O_{n+1}$ has a  C-$n$LHVM (\ref{C-n-PLHVM}) if and only if it has a D-$n$LHVM (\ref{D-n-PLHVM}).}

An $n$-local PT ${\bf{P}}=\Lbrack P({\bf{a}},b)\Rbrack$  over $O_{n+1}$  is said to be Bell local if the CT $\Lbrack P({\bf{a}},b|{\bf{x}},y)\Rbrack:=\Lbrack P({\bf{a}},b)\Rbrack$  over $\Delta_{n+1}$ with $m_i=m_B=1(i=1,2,\ldots,n)$ is Bell local, i.e.,
it can be written as
\Bq{BLP}
P({\bf{a}},b)=\sum_{\lm\in O_{n+1}}q(\lm)\prod_{i=1}^n
 P_i(a_i|\lm)\times P_B(b|\lm),\Eq
where $q(\lm), P_i(a_i|\lm), P_B(b|\lm)$ are PDs of $\lm,a_i$ and $b$, respectively. Indeed, every PT ${\bf{P}}=\Lbrack P({\bf{a}},b)\Rbrack$  over $O_{n+1}$ is Bell local since  it can be written as (\ref{BLP})
where $P_i(a_i|\lm)=\delta_{a_i|\lm_i},P_B(b|\lm)=\delta_{b,\lm_{n+1}}$ and
 $$\lm=(\lm_1,\ldots,\lm_n,\lm_{n+1}),q(\lm)=P(\lm_1,\ldots,\lm_n,\lm_{n+1}).$$
Using (\ref{lnL}) yields that
 \Bq{PlnL}
 \mathcal{PT}(O_{n+1})=
 \mathcal{PT}^{\textrm{Bell-local}}(O_{n+1})={\rm{conv}}(\mathcal{PT}^{n\textrm{-local}}(O_{n+1})).\Eq

{\bf{Corollary 5.2.}} {\it The set} $\mathcal{PT}^{n\textrm{-local}}(O_{n+1})$ {\it is  compact in the Hilbert space $\mathcal{T}(O_{n+1})$.}

From Definition 5.1, we observe  that when
${\bf{P}}\in\mathcal{PT}^{n\textrm{-local}}(O_{n+1})$, the marginal distributions satisfy
${\bf{P}}_{A_1\ldots A_n}={\bf{P}}_{A_1}\OO\ldots\OO{\bf{P}}_{A_n},
$
i.e.,
$${{P}}_{A_1\ldots A_n}(a_1,\ldots,a_n)={{P}}_{A_1}(a_1)\ldots{{P}}_{A_n}(a_n),\ \forall a_i\in[o_i];$$
especially,
\Bq{PP-product2}
{\bf{P}}_{A_1A_2}={\bf{P}}_{A_1}\OO{\bf{P}}_{A_2}.
\Eq
It is easy to construct a PT ${\bf{P}}$ that has no the property  (\ref{PP-product2}) and then is not $n$-local. This shows that not all Bell local PTs are $n$-local. By noticing that a convex combination of two product PTs is not necessarily a product PT, we conclude from the property  (\ref{PP-product2}) that  the set $\mathcal{PT}^{n\textrm{-local}}(O_{n+1})$ is not convex. However, it has many subsets that are star-convex.

Next, let us discuss the weak star-convexity of the set $\mathcal{PT}^{n\textrm{-local}}(O_{n+1})$ by finding star-convex subsets of it. To do so, for a fixed $1\le k\le n$, put
$${\widehat{A_k}}=A_1\ldots A_{k-1}A_{k+1}\ldots A_{n},{\bf{a}}_k=a_1\ldots a_{k-1}a_{k+1}\ldots a_n,$$
and take an $n-1$-partite product PT ${\bf{E}}_k=\Lbrack E_k({\bf{a}}_k)\Rbrack$ over $\prod_{i\ne k}[o_i]$  with
$E_k({\bf{a}}_k)=\prod_{i\ne k}Q_i(a_i),$ and define a set
\Bq{pCT-Ek}
\mathcal{CT}_{{\bf{E}}_k}^{n\textrm{-local}}(O_{n+1})=
\left\{{\bf{P}}\in \mathcal{PT}^{n\textrm{-local}}(O_{n+1}): {\bf{P}}_{\widehat{A_k}}={\bf{E}}_k\right\},
\Eq
where ${\bf{P}}_{\widehat{A_k}}$ denotes the marginal distribution of ${\bf{P}}$ on the subsystem $\widehat{A_k}$, i.e.,
${{P}}_{\widehat{A_k}}({\bf{a}}_k)=\sum_{a_k,b}P({\bf{a}},b).$
Define $Q_k(a_k)\equiv \frac{1}{o_k}$ and ${\bf{S}}_k=\Lbrack S_k({\bf{a}},b)\Rbrack$ by
$S_k({\bf{a}},b)=\prod_{i=1}^nQ_i(a_i)
\times\frac{1}{o_B},$
then ${\bf{S}}_k\in \mathcal{PT}_{{\bf{E}}_k}^{n\textrm{-local}}(O_{n+1})$ (Remark 5.1).

{\bf{Corollary  5.3.}} {\it  For any $k=1,2,\ldots,n,$ the set} $\mathcal{PT}_{{\bf{E}}_k}^{n\textrm{-local}}(O_{n+1})$  {\it is star-convex with a sun ${\bf{S}}_k$, i.e.,}
$$(1-t){\bf{S}}_k+t \mathcal{PT}_{{\bf{E}}_k}^{n\textrm{-local}}(O_{n+1})
\subset\mathcal{PT}_{{\bf{E}}_k}^{n\textrm{-local}}(O_{n+1}),\ \forall t\in[0,1].$$

{\bf{Corollary 5.4.}} {\it The set} $\mathcal{PT}^{n\textrm{-local}}(O_{n+1})$ {\it is path-connected.}

\section{Conclusions}

In this work, we have discussed the bilocality  of tripartite correlation tensors (CTs)  ${\bf{P}}=\Lbrack P(abc|xyz)\Rbrack$ over $\Delta_3=[o_A]\times[o_B]\times[o_C]\times[m_A]\times[m_B]\times[m_C]$ and probability tensors  (PTs)  ${\bf{P}}=\Lbrack P(a,b,c)\Rbrack$ over $O_3=[o_A]\times[o_B]\times[o_C]$, respectively, as well as the $n$-locality of $n+1$-partite CTs ${\bf{P}}=\Lbrack P({\bf{a}}b|{\bf{x}}y)\Rbrack$ over $\Delta_{n+1}=\prod_{i=1}^n[o_i]\times[o_B]\times\prod_{i=1}^n[m_i]\times[m_B]$  and PTs ${\bf{P}}=\Lbrack P({\bf{a}}b)\Rbrack$ over $O_{n+1}=\prod_{i=1}^n[o_i]\times[o_B]$, where $[n]$ stands for the set consisting  of $1,2,\ldots,n$. Based on a convex-decomposition lemma on a measurable function row-stochastic matrix, we have established a series of characterizations and properties of bilocal and $n$-local CTs and PTs, and obtained the following conclusions.

(1) Integration and summation descriptions (which we named C-biLHVM and D-biLHVM) for bilocality of a tripartite CT and PT are equivalent, denoted by C-biLHVM=D-biLHVM.

(2) A tripartite CT (resp. PT) ${\bf{P}}$ is  bilocal if and only if it is ``separable quantum", i.e., it can be generated by a pair of separable shared states together with a set of local POVMs (resp. a local POVM).

(3) The set $\mathcal{CT}^{\textrm{Bell-local}}(\Delta_3)$ of tripartite Bell local CTs with the same size $\Delta_3$ is just the convex hull of the set $\mathcal{CT}^{\textrm{bilocal}}(\Delta_3)$ of all bilocal CTs over  $\Delta_3$, while the convex hull of the set $\mathcal{PT}^{\textrm{bilocal}}(\Delta_3)$ of all bilocal PTs over  $O_3$ is just the set of all PTs  over  $O_3$.

(4) The set $\mathcal{CT}^{\textrm{bilocal}}(\Delta_3)$ forms a compact path-connected set w.r.t. the pointwise convergent topology on the index set $\Delta_3$ and has many star-convex subsets:
$$\mathcal{CT}_{A-{\bf{E}}}^{\textrm{bilocal}}(\Delta_3):=
\left\{{\bf{P}}\in \mathcal{CT}^{\textrm{bilocal}}(\Delta_3):\ {\bf{P}}_A={\bf{E}}\right\},\ \forall {\bf{E}}\in\mathcal{CT}([o_A]\times[m_A]),$$
$$\mathcal{CT}_{C-{\bf{F}}}^{\textrm{bilocal}}(\Delta_3):=
\left\{{\bf{P}}\in \mathcal{CT}^{\textrm{bilocal}}(\Delta_3):\ {\bf{P}}_C={\bf{F}}\right\},\ \forall {\bf{F}}\in\mathcal{CT}([o_C]\times[m_C]).$$

(5) Corresponding conclusions have been obtained for $n$-locality of $n+1$-partite CTs and PTs, including: C-$n$LHVM=D-$n$LHVM; ${\bf{P}}$ is in $\mathcal{CT}^{n\textrm{-local}}(\Delta_{n+1})$   if and only if it is ``separable quantum"; $\mathcal{CT}^{\textrm{Bell-local}}(\Delta_{n+1})$ is the convex hull of $\mathcal{CT}^{n\textrm{-local}}(\Delta_{n+1})$; $\mathcal{CT}^{n\textrm{-local}}(\Delta_{n+1})$ is a compact path-connected set w.r.t. the pointwise convergent topology on  $\Delta_{n+1}$ and has many star-convex subsets:
$$\mathcal{CT}_{{\bf{E}}_k}^{n\textrm{-local}}(\Delta_{n+1}):=
\left\{{\bf{P}}\in \mathcal{CT}^{n\textrm{-local}}(\Delta_{n+1}): {\bf{P}}_{\widehat{A_k}}={\bf{E}}_k\right\}
$$
for all ${\bf{E}}_k\in\mathcal{CT}(\prod_{i\ne k}[o_i]\times\prod_{i\ne k}[m_i])$, where ${\bf{P}}_{\widehat{A_k}}$ denotes the marginal distribution of ${\bf{P}}$ on the subsystem $\prod_{i\ne k}A_i$.

It is remarkable to point out that the bilocality is not a special case that $n=2$ of the $n$-locality discussed here, but it is a special case where $n=2$ of the $n$-locality ($n\ge2$)  introduced in \cite{[21]}, described by the following C-LHVM:
\begin{eqnarray*}
P(a_1\ldots a_{n+1}|x_1\ldots x_{n+1})&=&\int{\rm{d}}\lm_1\cdots\int{\rm{d}}\lm_n
\prod_{i=1}^n\rho_i(\lm_i)\times P_1(a_1|x_1,\lm_1)\\
&&\times
\prod_{i=2}^{n}P_i(a_i|x_i,\lm_{i-1}\lm_i)\times P_n(a_{n+1}|x_{n+1},\lm_n).
\end{eqnarray*}
Indeed, the $n$-locality here is based on the star network configuration \cite{Tavakoli}, while that of the work \cite{[21]} is a direct generalization of the bilocal scenario considered in \cite{Branciard2010}.

\section*{\bf Acknowledgements} {This work was supported by the National Natural Science Foundation of China (Nos. 11871318, 12271325), the Fundamental Research Funds for the Central Universities (GK202103003, GK202107014) and  and the Special Plan for Young Top-notch Talent of Shaanxi Province (1503070117).}

\section*{Appendix}
{\bf The proof of Lemma 2.1.} Let $N=n^m$ and $R_k=[\delta_{j,J_k(i)}]$, where $J_k({k\in[N]})$ are all of the mappings from $[m]$ into $[n]$. Here, we use $\max^*\{a_1,a_2,\ldots,a_n\}$ to denote the first maximal number of  $a_1,a_2,\ldots,a_n$  and define
$B^{(0)}(\lm)=[b_{ij}^{(0)}(\lm)]=B(\lm).$ For each $i\in[m]$, put
$$ b_{i,j^{(0)}_{i}(\lm)}^{(0)}(\lm)={\max}^*\left\{b_{i1}^{(0)}(\lm),b_{i2}^{(0)}(\lm),\ldots,b_{in}^{(0)}(\lm)\right\},$$ which is the first maximum value of entries in the $i$-th row of the matrix $B^{(0)}$, which is clearly $\Omega$-measurable w.r.t. $\lm$ on $\Lambda$. Define
$$
h_{ij}(\lm)=\left\{
\begin{array}{cc}
1, & j=j^{(0)}_{i}(\lm);\\
0, & {j\ne j^{(0)}_{i}(\lm)}
\end{array}
\right.\ R_{s_{1}(\lm)}=[h_{ij}(\lm)],
$$
$$\alpha_1(\lm)=\min\left\{b_{i,j^{(0)}_{i}(\lm)}^{(0)}(\lm)~|~i\in [m]\right\},$$$$ B^{(1)}(\lm)=[b^{(1)}_{ij}(\lm)]=B^{(0)}(\lm)-\alpha_1(\lm) R_{s_1(\lm)}.$$
Then $\alpha_1(\lm)$ is an entry of $B^{(0)}(\lm)=B(\lm)$ and $B^{(1)}(\lm)$ is a nonnegative matrix, and satisfies
$$B^{(0)}(\lm)=\alpha_1(\lm)R_{s_1(\lm)}+B^{(1)}(\lm),\ \  {\rm{zero}}(B^{(0)}(\lm))<{\rm{zero}}(B^{(1)}(\lm)),\eqno(A.1)$$
where ${\rm{zero}}(A)$ denotes the number of zero-entries of a matrix $A$.
Since $b_{ij}(\lm)$'s are $\Omega$-measurable w.r.t. $\lm$ on $\Lambda$, we see that $\alpha_1(\lm)$ and  entries of $B^{(1)}(\lm)$ are  $\Omega$-measurable  w.r.t. $\lm$ on $\Lambda$. Clearly, $R_{s_1(\lm)}$ is a $\{0,1\}$-RS matrix, depending on $\lm$.
To our aim, we have to replace it with a convex combination of $R_k$'s with $\Omega$-measurable coefficients. To do this, we put
$$ \Lambda^{(1)}_k=\{\lm\in\Lambda:R_{s_1(\lm)}=R_k\}(k=1,2,\ldots,N),\eqno{(A.2)}$$
then $\lm\in \Lambda^{(1)}_k$   if and only if $b_{ij^{(0)}_{i}(\lm)}^{(0)}(\lm)=b_{i,J_k(i)}^{(0)}(\lm)$ for all $i\in[m]$.
Hence, $\Lambda^{(1)}_k$ is an $\Omega$-measurable subset of $\Lambda$  and so its characteristic function $\chi_{\Lambda^{(1)}_k}$ is $\Omega$-measurable on $\Lambda$ for each $k\in[N]$. Thus, $c^{(1)}_k(\lm):=\chi_{\Lambda^{(1)}_k}(\lm)
\alpha_1(\lm)$ is nonnegative and $\Omega$-measurable w.r.t. $\lm$ on $\Lambda$. Since ${\Lambda^{(1)}_k}\cap{\Lambda^{(1)}_j}=\emptyset(k\ne j)$ and  $\cup_{k\in[N]}{\Lambda^{(1)}_k}=\Lambda$,  we see that
$$\sum_{k\in[N]}\chi_{\Lambda^{(1)}_k}(\lm)\alpha_1(\lm)=\alpha_1(\lm),\ \sum_{k\in[N]}\chi_{\Lambda^{(1)}_k}(\lm)R_k=R_{s_1(\lm)},\ \ \forall \lm\in\Lambda.\eqno(A.3)$$
Using Eqs. (A.1) and (A.3) yields that
$$
B^{(0)}(\lm)=\sum_{k\in[N]}c^{(1)}_k(\lm)R_{k}+B^{(1)}(\lm), \ \ \forall \lm\in\Lambda.
\eqno(A.4)$$
Similarly, we can decompose $B^{(1)}(\lm)$ in (A.4) as
$$
B^{(1)}(\lm)=\sum_{k\in[N]}c^{(2)}_k(\lm)R_{k}+B^{(2)}(\lm),\ \forall \lm\in\Lambda,
\eqno(A.5)$$
where $c^{(2)}_k(\lm)$ and $B^{(2)}(\lm)$ are  nonnegative and $\Omega$-measurable w.r.t. $\lm$ on $\Lambda$  with the property that ${\rm{zero}}(B^{(1)}(\lm))<{\rm{zero}}(B^{(2)}(\lm))$ for all $\lm\in\Lambda$. Thus,
$$
B^{(0)}(\lm)=\sum_{k\in[N]}c^{(1)}_k(\lm)R_{k}+\sum_{k\in[N]}c^{(2)}_k(\lm)R_{k}+B^{(2)}(\lm),\ \forall \lm\in\Lambda.
\eqno(A.6)$$
Continuously,   we can find nonnegative $\Omega$-measurable functions $c^{(1)}_k,c^{(2)}_k,\ldots,c^{(r)}_k$ and
nonnegative $\Omega$-measurable  matrices $B^{(1)},\ldots,B^{(r-1)},B^{(r)}$ such that
$$B^{(0)}(\lm)=\sum_{k\in[N]}\sum_{t=1}^{r-1}c^{(t)}_k(\lm)R_{k}+B^{(r)}(\lm),\ \ \forall \lm\in\Lambda,$$
with the property that
$${\rm{zero}}(B^{(0)}(\lm))<{\rm{zero}}(B^{(1)}(\lm))<{\rm{zero}}(B^{(2)}(\lm))
<\cdots<{\rm{zero}}(B^{(r-1)}(\lm))\le mn.$$
Thus, after doing at most $mn$ steps, we arrive at the case where $B^{(r)}=0$ and then obtain that
$$B(\lm)=B^{(0)}(\lm)=\sum_{k=1}^N\alpha_k(\lm)R_{k},\ \ \forall \lm\in\Lambda,$$
where the coefficient functions $\alpha_k=\sum_{t=1}^{r-1}c^{(t)}_k(k\in[N])$, which are nonnegative and $\Omega$-measurable functions on $\Lambda$.  The proof is completed.

\end{document}